\documentclass[11pt]{article}
\usepackage[a4paper,hmarginratio=1:1,vmarginratio=2:3,totalwidth=15.6cm,
totalheight=21.002cm]{geometry}
 \usepackage{verbatim} 
\usepackage{latexsym}
\renewcommand{\baselinestretch}{1.29}
\usepackage{amsmath}
\usepackage{amssymb}
\usepackage{amsfonts} 
\usepackage{epsfig}

\newcommand{\cD}{{\cal D}}

\newcommand{\cL}{{\cal L}}

\newcommand{\cO}{{\cal O}}

\newcommand{\cZ}{{\cal Z}}

\newcommand{\geqsim}{\,\raisebox{-0.6ex}{$\buildrel > \over \sim$}\,}

\renewcommand{\Im}{\text{Im}\ }

\newcommand{\ra}{\rightarrow}
\newcommand{\be}{\begin{equation}}
\newcommand{\ee}{\end{equation}}
\newcommand{\bea}{\begin{eqnarray}}
\newcommand{\eea}{\end{eqnarray}}

\DeclareMathSymbol{\mg}{\mathrel}{symbols}{"1D}

\long\def\symbolfootnote[#1]#2{\begingroup%
\def\thefootnote{\fnsymbol{footnote}}\footnote[#1]{#2}\endgroup} 

\newcounter{oldcounter}
\addtocounter{equation}{1}
\begin{document}

 \begin{flushright}
{\hfill{ CPHT-RR106.1009, LPT-Orsay 09-77}}\\
{CERN-PH-TH/2009-184}
\end{flushright}

\thispagestyle{empty}

\vspace{1.2cm}
\begin{center}
{\Large {\bf MSSM Higgs with dimension-six operators. 
}}

\vspace{0.2cm}
\end{center}

\begin{center}
{\bf I. Antoniadis$^{\,a, b\,}$,
E. Dudas$^{\,b, c\,}$,
D.~M. Ghilencea$^{\,a,\,b,\,}$\symbolfootnote[2]{
on leave from Theoretical Physics Department,
IFIN-HH Bucharest, POBox MG-6, Romania.}, 
P. Tziveloglou$^{\,a,\,d,\,}$\footnote{
E-mail addresses:\,\, Ignatios.Antoniadis@cern.ch,
Emilian.Dudas@cpht.polytechnique.fr,

$\,\,\,\,$Dumitru.Ghilencea@cern.ch,
PT88@cornell.edu}}\\

 \vspace{0.2cm}
 {\it $^a $Department of Physics, CERN - Theory Division, 1211 Geneva 23,
 Switzerland.}\\[2pt]
 {\it $^b $Centre de Physique Th\'eorique, Ecole Polytechnique, CNRS, 91128
   Palaiseau, France.}\\[2pt]
 {\it $^c $LPT, UMR du CNRS 8627, B\^at 210, Universit\'e de Paris-Sud,
 91405 Orsay Cedex, France.}
 \\[2pt]
 {\it $^d $ Department of Physics, Cornell University, Ithaca, NY 14853 USA.}
 \end{center}

\def\baselinestretch{1.1}
\begin{abstract}
\noindent 
We investigate an extension of the MSSM Higgs sector by
including the effects of all dimension-five and  dimension-six 
effective operators  and their  associated supersymmetry breaking
terms. The corrections to the masses of the
neutral CP-even and CP-odd Higgs bosons
due to the $d\!=\!5$ and $d\!=\!~6$ operators are computed. 
When the $d\!=\!5$ and $d\!=\!6$ operators are generated by the same physics
(i.e. when suppressed by powers of the same scale $M$),
due to the relative $\tan\beta$ enhancement of the latter, 
which compensates  their extra scale suppression ($1/M$), 
the mass corrections from  $d=6$ operators 
can be comparable to those of $d=5$ operators, 
even for conservative values of the scale $M$.
We identify the effective operators with the largest 
individual corrections  to the lightest Higgs mass
and discuss whether at the microscopic level and in the simplest
cases, these operators are generated by ``new physics'' 
with a sign consistent with an increase of $m_h$. 
Simple numerical estimates easily allow  an increase of $m_h$ due to $d=6$
operators alone in the region of $10-30$ GeV, while for a much larger
increase light new states beyond MSSM may be needed, 
in which case the effective description is unreliable. 
  Special attention is paid to the treatment of the 
  effective operators with higher derivatives.
  These can be removed by non-linear field redefinitions or by an
  ``unfolding'' technique, which effectively ensure that any ghost
  degrees of freedom (of mass $\geqsim M$) are integrated out 
  and absent in the effective theory at scales much smaller than $M$.
Considering general coefficients of the 
susy operators with a scale of new physics above the LHC reach, 
it is possible to increase the tree-level prediction for the Higgs
mass to the LEPII bound, thus 
alleviating  the MSSM fine-tuning.
\end{abstract}

\newpage
\setcounter{page}{1}
\setcounter{footnote}{0}
\tableofcontents

\section{Introduction}

The coming LHC experiments are a great opportunity to test
directly  the  old idea of low-energy supersymmetry, as a 
possibility of new physics
beyond the Standard Model. In the minimal supersymmetric version of this
model (MSSM) one obtains definite predictions in particular for the 
Higgs sector. For an agreement with 
the LEPII  constraints for the mass of the SM-like
Higgs of $m_h\!>\!114.4$ GeV \cite{higgsboundLEP}, 
 the MSSM requires that quantum corrections 
lift its tree-level bound  $m_h\!\leq\! m_Z$. This is indeed possible
 and acceptable within the allowed parameter space, increasing 
the Higgs mass above the LEPII bound (for a recent MSSM
fit see \cite{AbdusSalam:2009qd}). However, larger quantum corrections
usually require larger soft terms, making it more difficult
to satisfy the electroweak constraint $v^2=-m_{soft}^2/\lambda$
with $\lambda$ the MSSM effective quartic coupling
and $m_{soft}^2$ a linear combination of soft (masses)$^2$. 
 With $\lambda$ fixed by the gauge sector and with 
$m_{soft}\!\sim\! TeV$ this condition  is more difficult to respect,
given the negative searches for supersymmetry so far
and the mass bounds for sparticles. 
As a result, the MSSM appears fine-tuned 
\cite{Barbieri:1998uv,Chankowski:1998xv,Chankowski:1997zh,Kane:1998im}
although there is no universally agreed fine-tuning measure or
exact value. For further discussion in this direction 
see \cite{Giudice:2006sn}.  
This situation could even be seen as undermining
the original motivation for supersymmetry, prompting  
alternatives such as~\cite{Giudice:2007fh}.

If one maintains the idea of TeV-scale supersymmetry, 
such a problem  of the MSSM must be addressed.
The most common idea to solve it is to assume that {\it new physics} 
beyond the MSSM is present somewhere in the region of a few TeV. 
To investigate this possibility,
 a general, model-independent approach can be considered, 
by parametrising this new physics using effective operators.
This is possible by organising such
operators in inverse powers of the scale $M$ of new physics which,
when integrated out, generates these effective operators.
One can later address the question of 
what ``new physics'' may generate these operators.
In \cite{Dine} operators
of dimension $d\!=\!5$ were considered in the Higgs sector, 
together with their microscopic 
origin and implications for the Higgs mass ($m_h$). 
Further analysis  including all baryon 
and lepton
number conserving $d\!=\!5$ operators beyond MSSM  was done in 
\cite{ADGT,Antoniadis:2008ur}, 
showing how generalised, spurion dependent field redefinitions 
reduce the number of effective operators to an irreducible, 
minimal set. As a result, the number of independent parameters 
is reduced with the benefit of improving the predictive power 
of the method. 
Further analysis of the MSSM with ($d\!=\!5$) effective operators
 studied  the stability of the Higgs potential
with these operators \cite{Blum:2009na},
the effects on the neutralino sector \cite{Berg:2009mq}, baryogenesis
\cite{Blum:2008ym}, CP violation~\cite{Ham:2009gu}
or fine-tuning \cite{Cassel:2009ps}. 
The presence of these operators of $d=5$ can increase the 
effective quartic coupling $\lambda$ of the Higgs field and as a result
the fine tuning \cite{r3} for the MSSM electroweak scale is reduced
 \cite{Cassel:2009ps}  (see also \cite{Batra:2003nj}). 
One obtains one-loop values $114\!\leq\! m_h\!\leq 130$ 
GeV with a very acceptable  fine-tuning
$\Delta\!\leq\! 10$ at one-loop, for $M\!\sim\!8$ to $10$ TeV
\cite{Cassel:2009ps}.

The purpose of this work is to extend these studies 
by  considering,  in a systematic way,
operators of dimension $d=6$ that can account for new physics
beyond  the MSSM Higgs sector. The motivation is  that such operators
can bring relevant contributions to $m_h$, even in the absence of
$d=5$ operators.
This is indeed possible, since not all $d\!=\!6$ 
operators are necessarily generated by the same 
new physics as the $d=5$ ones. Even if both $d=5$ and $d=6$ operators
are present, they could also be suppressed by a 
different high scale, if generated by different new physics; 
this possibility can, in principle, also
be read from our results by keeping track of 
their coefficients. 
Finally,
if all or some of the $d=5,6$ operators are generated by same new physics,
while suppressed by an extra $1/M$ factor relative to the 
leading $d\!=\!5$ operators, the $d=6$ operators
can nevertheless 
have an impact for the large $\tan\beta$ region of the
parameter space. Indeed,  some $d\!=\!6$ operators acquire 
an  enhancement factor  ($\tan\beta$) relative to the
 $d\!=\!5$ operators, which compensates for their extra scale suppression.
As a result, their effects on the Higgs mass can be
comparable to those of $d\!=\!5$ operators and it is interesting
to examine, in this setup, the new corrections to $m_h$
at classical level.  The study is also
relevant for examining the limits of the approximation of expanding
in powers of $1/M$ by comparing leading 
 and sub-leading 
terms of this expansion. Such corrections to 
$m_h$ can be as large as loop corrections to $m_h$.
We identify {\it individual} $d=6$ operators with 
the largest correction to $m_h$, 
and discuss their possible microscopic origin
and the signs they are generated with in the simplest cases.
Although we do not provide detailed examples
of high-energy physics that give the desired signs, 
considering general coefficients of the susy operators of 
$d=6$, 
our results show that 
one can increase the tree-level prediction for the Higgs 
mass to the LEPII bound
(alleviating the fine-tuning of the MSSM
as noticed earlier for $d=5$ operators \cite{Dine,Cassel:2009ps}),
even for a scale of new physics above the LHC reach.

While studying higher dimensional
operators, one problem is associated with the presence in some of these
of higher derivatives, i.e. the presence of ghosts degrees
of freedom in the spectrum. In some cases one
can use the equations of motion  to set these operators ``on-shell''
\cite{Georgi:1991ch,Politzer:1980me,Arzt}, and remove the extra
derivatives. We investigate this procedure and show that this
ultimately means integrating out the ghost degrees of freedom.  
While this ``on-shell'' method is
correct in the leading order (in $1/M$), it is not 
true beyond it.  Appendix~\ref{appendixB} provides detailed
examples which investigate these issues, and  supports this 
statement (see in particular Appendix~\ref{appendixB2}).
A more general and correct procedure is to use instead non-linear field
redefinitions  to remove the derivative operators.
A third and more interesting method is to re-write (``unfold'')
the original theory
with higher derivatives as a second-order theory
(i.e. with at most two derivatives) with additional
(ghost) superfields of mass of order $M$ \cite{ADG}. 
After integrating classically these fields one obtains in the
low energy action below $M$, an effective theory without 
higher derivatives and with (classically) renormalised interactions. 
The results  obtained are identical to those obtained by using the non-linear
field redefinitions mentioned earlier; in the leading
order in $1/M$ the method of setting ``on-shell'' the 
operators with extra derivatives by equations of motion gives 
similar results.

The aforementioned
 presence of the ghost degrees of freedom near the scale $M$ simply
warns us that beyond this scale the theory is unstable and UV
incomplete. This is a generic situation in all effective theories, even
in those obtained from renormalisable ones by integrating out 
a massive state and after truncating the effective action
to a given order (in $1/M$).
These problems are also present in our discussion with $d\!=\!6$ operators.
In the end, one  eliminates the $d\!=\!5$ operators with extra derivatives 
via field redefinitions,  to leave only polynomial (in superfields)
$d\!=\!5$ operators, and $d\!=\!6$ operators in 
which it is possible to use the equations of motion (with a
similar result with integrating out the ghost degrees of freedom).
In the presence of supersymmetry breaking  additional effects are
present, like: $\mu$-term renormalisation by susy-breaking terms, 
soft terms renormalisation, discussed in detail
in  \cite{ADGT}.

The plan of the paper is as follows: Section~\ref{section2} presents
the list of operators and clarifies which of them have relevant
contributions to the scalar potential. In Section~\ref{section3} the scalar
potential of the Higgs in MSSM with $d\!=\!5,6$ operators  is computed.
Section~\ref{section4} shows the results for the 
masses of the CP even/odd higgses. We identify the operators 
with the largest contribution in Section~\ref{section5}.
The Appendix provides technical details 
and shows how to replace higher derivative operators 
by non-derivative ones in an effective action with $d=5, 6$
operators.

\bigskip\medskip
\section{MSSM Higgs sector with d=5 and d=6 operators}\label{section2}

The relevant part of the Lagrangian of our model contains a piece
$\cL_0$ of the MSSM higgs sector, together with that due
to relevant $d=5$ and $d=6$  operators. 
For $\cL_0$  we have
\medskip\bea\label{lo}
\cL_0=\int d^4\theta \,\,\sum_{i=1,2}
Z_i(S,S^\dagger)\,H_i^\dagger\,e^{V_i}\,H_i
+\bigg\{\int d^2\theta \,\,\mu_0\,(1+B_0\,m_0
\,\,\theta\theta)\,H_1.H_2+h.c.\bigg\}
\eea
in a standard notation. Here
$Z_{i}(S,S^\dagger)=1-c_{i}\,m_0^2\,\theta\theta\overline\theta\overline\theta$
with $i=1,2$ and $c_i=\cO(1)$,
 $m_0$ is the supersymmetry breaking scale in the visible sector,
 $m_0=\langle F_{hidden}\rangle/ M_{Planck}$ with $F_{hidden}$ an
auxiliary field in the hidden sector. As usual we assume this breaking
is transmitted to the visible sector through gravitational
interactions mediated by $M_{Planck}$.

We extend this Lagrangian
by $d=5$ and $d=6$  operators. In the first class we have
\medskip  
\begin{eqnarray} 
\mathcal{L}_{1} &=&\!\!\frac{1}{M}\int d^{2}\theta \,
\,\zeta(S)\,(H_{2}.H_{1})^{2}\!+\!h.c.
=2\,\zeta_{10}\,(h_2.h_1)(h_2.F_1+F_2.h_1)+\zeta_{11}\,m_0\,(h_2.h_1)^2+h.c, 
\nonumber\\
\mathcal{L}_{2} &=&\!\!\frac{1}{M}\int d^{4}\theta \,\,\Big\{%
\,A(S,S^{\dagger })D^{\alpha }\Big[B(S,S^{\dagger })\,H_{2}\,e^{-V_{1}}\Big]%
D_{\alpha }\Big[C(S,S^{\dagger })\,e^{V_{1}}\,H_{1}\Big] 
+h.c.\Big\} \label{dimensionfive}
\end{eqnarray}
where\footnote{Other notations used: in
 \cite{ADGT} $\eta_2\!=\!2\zeta_{10}\mu_0^*$, 
$\eta_3\!=\!-2\,m_0 \zeta_{11}$; in \cite{Cassel:2009ps}
$\eta_2\!\ra\! \zeta_1$, $\eta_3\!\ra\! \zeta_2$; in \cite{Dine}
$\eta_2\!\ra\! 2\epsilon_{1r}$, $\eta_3\!\ra\! 2\epsilon_{2r}$}
\bea
\frac{1}{M}\,\zeta(S)
= \zeta_{10}+ \zeta_{11}\,m_0\,\theta\theta,\,\,
\qquad \zeta_{10},\, \zeta_{11}\sim 1/M,
\eea

\medskip\noindent
with $S$ the spurion superfield, $S=\theta\theta m_0$.
We assume that 
\bea
m_0\ll M
\eea
 so that the  effective theory approach is reliable.
If this condition is not respected and the ``new physics'' is represented by
``light'' states (like the MSSM states), 
then one should work in the model where these 
are not integrated out. 
$A, B, C$ are general functions, which take into account supersymmetry
breaking associated with these operators
so, for example:
\bea
A(S,S^\dagger)=
a_0+a_1\,S+a_1^*\, S^\dagger
+a_2\,S\,S^\dagger, \qquad (\textrm{similar\,\,for\, \,B,C})
\eea
They are general and account for effects of supersymmetry breaking in the
presence of some massive states which when integrated out generate
 $\cL_{1,2}$ with these susy breaking terms.

$\cL_2$ is eliminated by generalised, spurion-dependent
field redefinitions as it was  showed in detail in \cite{ADGT}.
We assume this procedure was already implemented, therefore
only $\cL_1$ is relevant for the discussion below.
These redefinitions  bring however a renormalisation
of the usual MSSM soft terms and of the $\mu$ term, 
and additional corrections of order $1/M^2$. The latter are
 corrections to the $d=6$ operators that are relevant 
for the Higgs sector, that we present shortly.
Since we shall write down {\it all} $d=6$ operators, these
corrections are then ultimately accounted for by
renormalisations (redefinitions) of the coefficients of
the $d=6$ terms. Since we take these coefficients arbitrary,
without any restriction to generality
we can assume these redefinitions are  already implemented.

The list of $d=6$ operators is \cite{Piriz:1997id}
\bea
\mathcal{O}_{j} &=&
\frac{1}{M^{2}}\int d^{4}\theta \,\,
\mathcal{Z}_{j}(S,S^{\dagger })\,\,
(H_{j}^{\dagger }\,e^{V_{j}}\,H_{j})^{2}, \quad j\equiv 1,2. 
\nonumber\\[-2pt]
\mathcal{O}_{3} &=&
\frac{1}{M^{2}}\int d^{4}\theta \,\,\mathcal{Z} _{3}(S,S^{\dagger
})\,\,
(H_{1}^{\dagger 
}\,e^{V_{1}}\,H_{1})\,(H_{2}^{\dagger }\,e^{V_{2}}\,H_{2}),
\nonumber\\
\mathcal{O}_{4} &=&\frac{1}{M^{2}}\int d^{4}\theta \,\,
\mathcal{Z}_{4}(S,S^{\dagger })\,\,(H_{2}.\,H_{1})\,(H_{2}.\,H_{1})^{\dagger },
\nonumber\\
\mathcal{O}_{5} &=&\frac{1}{M^{2}}\int d^{4}\theta \,\,\mathcal{Z}%
_{5}(S,S^{\dagger })\,\,(H_{1}^{\dagger 
}\,e^{V_{1}}\,H_{1})\,\,H_{2}.\,H_{1}+h.c.
\nonumber\\
\mathcal{O}_{6} &=&
\frac{1}{M^{2}}\int d^{4}\theta \,\,\mathcal{Z}_{6}(S,S^{\dagger
})\,\,
(H_{2}^{\dagger }\,e^{V_{2}}\,H_{2})\,\,H_{2}.\,H_{1}+h.c. 
\nonumber\\
\mathcal{O}_{7} &=&\frac{1}{M^{2}}\int d^{2}\theta \,\,
\mathcal{Z}_{7}(S,0)\,\frac{1}{16\,g^2\,\kappa}\,{\rm Tr}\,
W^{\alpha }\,W_{\alpha }\,(H_{2}\,H_{1})+h.c.
\nonumber\\
\mathcal{O}_{8} &=&\frac{1}{M^{2}}\int d^{4}\theta
 \,\,\Big[\mathcal{Z}_{8}(S,S^{\dagger
   })\,\,(H_{2}\,H_{1})^{2}+h.c.\Big]
\label{operators18}\eea

\bigskip\noindent
where $W^\alpha=(-1/4)\,\overline D^2 e^{-V} D^\alpha\, e^V$
is the chiral field strength
of $SU(2)_L$ or $U(1)_Y$ vector superfields $V_w$ and $V_Y$ respectively.
 Also  $V_{1,2}=V_w^a
(\sigma^a/2)+(\mp 1/2)\,V_Y$ with the upper (minus) sign for $V_1$.
The expressions of these operators in component form,
are given in Appendix~\ref{appendixA}.
The remaining $d=6$ operators are:
\bea
\mathcal{O}_{9} &=&\frac{1}{M^{2}}\int d^{4}\theta \,\,
\mathcal{Z}_{9}(S,S^{\dagger })\,\,H_{1}^{\dagger }\,
\overline{\nabla}^{2}\,e^{V_{1}}\,\nabla ^{2}\,H_{1}  \nonumber \\[-2pt]
\mathcal{O}_{10} &=&\frac{1}{M^{2}}\int d^{4}\theta \,\,
\mathcal{Z}_{10}(S,S^{\dagger })\,\,H_{2}^{\dagger }\,\overline{\nabla } 
^{2}\,e^{V_{2}}\,\nabla ^{2}\,H_{2}  \nonumber \\
\mathcal{O}_{11} &=&\frac{1}{M^{2}}\int d^{4}\theta \,\,\mathcal{Z}
_{11}(S,S^{\dagger })\,\,H_{1}^{\dagger }\,e^{V_{1}}\,\nabla ^{\alpha 
}\,W_{\alpha }^{(1)}\,H_{1}  \nonumber \\
\mathcal{O}_{12} &=&\frac{1}{M^{2}}\int d^{4}\theta \,\,\mathcal{Z} 
_{12}(S,S^{\dagger })\,\,H_{2}^{\dagger }\,e^{V_{2}}\,\nabla ^{\alpha 
}\,W_{\alpha }^{(2)}\,H_{2}  \label{der0} 
 \nonumber \\ 
\mathcal{O}_{13} &=&\frac{1}{M^{2}}\int d^{4}\theta \,\,\mathcal{Z}
 _{13}(S,S^{\dagger })\,\,H_{1}^{\dagger }\,e^{V_{1}}\,
 \,W_{\alpha }^{(1)}\,\nabla^\alpha\,H_{1}  \nonumber \\ 
\mathcal{O}_{14} &=&\frac{1}{M^{2}}\int d^{4}\theta \,\,\mathcal{Z} 
 _{14}(S,S^{\dagger })\,\,H_{2}^{\dagger }\,e^{V_{2}}\,
 \,W_{\alpha }^{(2)}\,\nabla^\alpha\,H_{2}  \label{der} 
\eea
Also
$\nabla_{\alpha }\,H_{i}=e^{-V_{i}}\,D_{\alpha }\,e^{V_{i}}
 H_i$ and  $W_\alpha^i$ is the field strength of $V_i$.
To be even more general, in the above operators
 one should actually include spurion dependence
under any $\nabla_\alpha$, 
of arbitrary coefficients to include supersymmetry 
breaking  effects associated to them. 
Finally, the wavefunction coefficients introduced above 
have the structure
\medskip
\bea
\frac{1}{M^2}\,\cZ_i(S,S^\dagger)=\alpha_{i0}
+\alpha_{i1}\,m_0\,\theta\theta
+\alpha_{i1}^*\,m_0\,\overline\theta\overline\theta
+\alpha_{i2}\,m_0^2\,\theta\theta\overline\theta\overline\theta,\qquad
\alpha_{ij}\sim 1/M^2.
\eea

\medskip
Regarding the origin of these operators: $\cO_{1,2,3}$ can be
 generated in MSSM with an additional,
 massive $U(1)'$ gauge boson or $SU(2)$ triplets integrated out \cite{Dine}.
$\cO_4$ can be generated by a massive gauge singlet or $SU(2)$
 triplet, 
while $\cO_{5,6}$
can be generated by a combination of $SU(2)$ doublets
and massive gauge singlet. $\cO_7$ is essentially a 
threshold correction to the gauge coupling, with a 
moduli field replaced by the Higgs.  $\cO_8$ 
exists only in non-susy case, but is generated when redefining away
the $d=5$ derivative operator \cite{ADGT}, thus we keep it.

Let us consider for a moment the operators $\cO_{9,...14}$ in
the exact supersymmetry case. Then, we can set ``on-shell'' some of  
these, by using the eqs of  motion\footnote{
Superpotential convention: $\int d^2\theta\mu_0\,H_1.H_2=
\int d^2\theta\,\mu_0
 \,H_1^T\, (i\sigma_2)\,H_2\equiv \int d^2\theta\, 
 \mu_0\,\epsilon^{ij}\,H_1^i\,H_2^j$;
 \,\,$\epsilon^{12}=1=-\epsilon^{21}$.}:
\medskip
\bea\label{hdo2}
&& -\frac{1}{4}\,\overline D^2\,(H_2^\dagger \,e^{V_2})
+\mu_0\,H_1^T\,(i\sigma_2)=0,\qquad
\frac{1}{4}\,\overline D^2\,(H_1^\dagger \,e^{V_1})
+\mu_0\,H_2^T (i\sigma_2)=0
\eea

\medskip\noindent
With this we find that in the supersymmetric case
\footnote{Also using $(i\sigma_2)\,e^{-\Lambda}=
e^{\Lambda^T}\,(i\sigma_2)$;\, $\Lambda\equiv \Lambda^a\,T^a$;\,
 $(i\sigma_2)^T =-( i\sigma_2)$;\, $(i\sigma_2)^2=-1_2$}:
\medskip
\bea\label{hdo1}
\cO_{9}\sim \int d^4\theta\,\,
H_1^\dagger \,\overline \nabla^2\, e^{V_1}\,\nabla^2\,H_1
=16\,\vert\mu_0\vert^2\,\int d^4\theta\,H_1^\dagger\,e^{V_1}\,H_1.
\eea

\medskip\noindent
and similar for $\cO_{10}$. 
Regarding $\cO_{11,12}$, in the  supersymmetric
case they vanish, following the definition of $\nabla^\alpha$ and
an integration by parts. 
Further, $\cO_{13,14}$ are similar to $\cO_{9,10}$,
which can be seen by using the definition of $W_\alpha^{(i)}$ 
and the relation between  $\nabla^2$, ($\overline\nabla^2$) and
$D^2$, ($\overline D^2$).

In conclusion, in the exact supersymmetric case, $\cO_{9...14}$ give at 
most wavefunction renormalisations of operators already included.
This was shown by using the equations of motion 
(``on-shell'' method -- we return to this issue shortly).
Let us now consider supersymmetry breaking associated to these
operators, due to their spurion dependence.
Turning on supersymmetry breaking should not bring physical
effects, as showed explicitly in \cite{ADGT} and could only give
soft terms and $\mu$-term renormalisation by $\cO(1/M^2)$
corrections. 
Since these terms are anyway renormalised by $\cO_{1,...8}$, 
where  spurion dependence is included with {\it arbitrary}
coefficients,  then there is no loss of generality to ignore the
supersymmetry breaking effects associated to
$\cO_{9,...14}$ in the following discussion,
which are anyway taken into account by $\cO_{1,..8}$.
Following this discussion, one concludes that
$\cO_{9,...,14}$ are not relevant for the analysis
of the Higgs potential performed below.
Finally, there  can be an additional operator of $d=6$
from the gauge sector, $\cO_{15}=(1/M^2)\int d^2\theta
\,\,W^\alpha\Box W_\alpha$ which could affect 
the Higgs potential\footnote{Its complete gauge invariant form 
is $\int d^4 \theta \ Tr \ e^V W^{\alpha} e^{-V} D^2 (e^V
  W_{\alpha}  e^{-V})$.}.  Using the equations of motion for the
gauge field it can be shown that $\cO_{15}$ gives a renormalisation 
of $\cO_{1,2,3}$, so its effects are ultimately included,
since the coefficients $\cZ_{1,2,3}$ are arbitrary.

The careful reader may question the above use of the eqs of motion 
in some of the higher dimensional operators, in order to
essentially remove  those with more than two  
derivatives ($\cO_{9,..,15}$). This
 ``on-shell''procedure is justified by previous works 
\cite{Georgi:1991ch,Politzer:1980me} and further detailed in \cite{Arzt}.
A more general and correct approach is to 
use instead non-linear field redefinitions\footnote{These 
are actually employed to prove  this ``on-shell'' method
\cite{Arzt}.}
or an ``unfolding'' technique (see later).
These two generally valid
approaches are discussed in detail  in Appendix~\ref{appendixB}.
We used these two approaches  to check the validity of the
above ``on-shell'' procedure, for the cases and approximation
in which we applied it.
This was also done  to clarify, from a general
perspective,  what actually means to set  ``on-shell'' the higher 
derivative operators.

To this purpose, consider for simplicity the case of operator $\cO_9$
without gauge fields, 
when $\cO_9\sim (1/M^2)\,\int d^4\theta \,\,H_1^\dagger \Box H_1$.
A Lagrangian with such a higher derivative 
operator contains additional poles corresponding
to ghosts degrees of freedom.
As shown in \cite{ADG}, see also  Appendix~\ref{appendixB1},
such theory can be reformulated and ``unfolded'' into
a second order one (i.e with no more than two derivatives)
with   (one or two) additional ghost superfields of 
mass of the order $M$.
In such an effective theory, at energies well below the scale
$M$, such ghost-like states can then be integrated out.
The result is a wavefunction renormalisation only, which is
in agreement with the result obtained by the ``on-shell'' method discussed
above. Therefore,   using the eqs of motion
 to set ``on-shell''  the higher derivative operator as done in
 (\ref{hdo2}), (\ref{hdo1}) corresponds to integrating out the
massive ghost degrees of freedom associated with such operator.
For  details see Appendix~\ref{appendixB1}.

A result similar to the ``unfolding'' method is obtained by 
using non-linear field redefinitions. This was detailed  in
Appendix~\ref{appendixB2} for the case of 
$d=5$ operators. There it is shown that in the leading order in $1/M$
the ``unfolding'' method (integrating out the ghosts), the
nonlinear field redefinition method  and the ``on-shell''
method give similar results. Beyond this $1/M$ order however, the 
``on-shell'' method should be appropriately modified to 
use the  Euler-Lagrange equations for a higher derivative Lagrangian.

 With these clarifications one can safely say that
 $\cO_{9,...,15}$ are not relevant for the following discussion
 of the Higgs potential.
In conclusion the  list of $d=6$ operators that remain for our study of 
the Higgs sector beyond MSSM is that of eq.(\ref{operators18}). 
Let us stress that not all the remaining operators $\cO_{1,..,8}$ of
(\ref{operators18})
are necessarily  present or generated in a detailed model. 
Symmetries and details of the ``new physics'' beyond the MSSM that
generated them,  may  forbid or favour the presence of some of them.
Therefore, we regard these remaining
operators as {\it  independent}  of each other, although in specific models
correlations may exist among their coefficients $\cZ_i$. 
It is  important to keep all these operators in the
analysis, for the purpose of identifying which of
them has the largest individual contribution to the Higgs mass,
which is one of the main interests of this analysis.
Finally,  some of the $d=6$ operators can in principle
be present even in the absence of the $d=5$ operators, if these classes of
operators are generated by integrating different ``new physics''.
In specific models one simply sets to zero, in the results
below, the coefficients of those operators of $d\!=\!5$ and/or $d\!=\!6$
not present/generated.

\section{The scalar potential with d=5 and d=6 operators}\label{section3}

Following the previous discussion, 
the overall Lagrangian of the model is
\medskip
\bea\label{LL}
\cL_H = \cL_0+\cL_1+\sum_{i=1}^{8}\,\cO_i
\eea

\medskip\noindent
with the MSSM higgs Lagrangian $\cL_0$ of eq.(\ref{lo}), $\cL_1$
of eq.(\ref{dimensionfive}) and $\cO_{1,2,....,8}$ of 
eq.(\ref{operators18}).

With the results in Appendix~\ref{appendixA} we find
the following contributions to the scalar potential:
\bea
V_F=\frac{\partial^2\,K}{\partial\,h_i\,\partial\, h_j^*}
\,F_i\,F_j^*=\vert F_1\vert^2+
\vert F_2\vert^2+
\frac{\partial^2\,K_6}{\partial\,h_i\,\partial \,h_j^*}
\,F_i\,F_j^*\label{VV}
\eea

\smallskip\noindent
where $K_6$ is the contribution of $\cO(1/M^2)$ to the K\"ahler 
potential due to  $\cO_{1,...8}$. The first two terms in the rhs give
($h_i$ denote $SU(2)_L$ doublets, $\vert h_i\vert^2\equiv h_i^\dagger \,h_i$):
\medskip
\bea
V_{F,1}&\equiv &\vert F_1\vert^2+
\vert F_2\vert^2
\nonumber\\
&=&\vert \mu_0+2\,\zeta_{10}\,h_1.h_2\vert^2\,\,
\big(\vert h_1\vert^2+\vert h_2\vert^2\big)
\nonumber\\[3pt]
&+&
\Big[\mu_0^*\,\Big(
\vert h_1\vert^2\,\rho_{21}+\vert h_2\vert^2\,\rho_{11}
+(h_1.h_2)^\dagger\,(\rho_{22}+\rho_{12})\Big)+h.c.\Big]
\label{VF1}
\eea

\medskip\noindent
obtained using (\ref{arhos}) and where $\rho_{ij}$ are functions of $h_{1,2}$:
\medskip
\bea\label{rhos0}
\rho_{11}
&=&
-(2\alpha_{10}\,\mu_0+\alpha_{40}\mu_0+\alpha_{51}^*\,m_0)\vert h_1\vert^2
-(\alpha_{30}\,\mu_0+\alpha_{40}\mu_0+\alpha_{61}^*\,m_0)\,\vert h_2\vert^2
\nonumber\\
&&-
(\alpha_{41}^*\,m_0+\alpha_{50}^*\,\mu_0)\,(h_2.h_1)^*+
\big[\,(\alpha_{60} +2\,\alpha_{50})\,\mu_0+2\alpha_{81}^*\,m_0\big]\,
(h_1.h_2)
\nonumber\\
\rho_{12}
&=&
\,\,\,(2\alpha_{11}^*\,m_0+\alpha_{50}^*\,\mu_0)\vert h_1\vert^2
+(\alpha_{31}^*\,m_0+\alpha_{50}^*\,\mu_0)\,\vert h_2\vert^2
\nonumber\\
&&-\big[(2\alpha_{10}+\alpha_{30})\,\mu_0+\alpha_{51}^*\,m_0\big]\,(h_1.h_2)
+\alpha_{51}^*\,m_0\,(h_2.h_1)^*
\eea
\bea
\rho_{21}
&=&
 -(2\alpha_{20}\,\mu_0+\alpha_{40}\mu_0+\alpha_{61}^*\,m_0)\vert h_2\vert^2
 -(\alpha_{30}\,\mu_0+\alpha_{40}\mu_0+\alpha_{51}^*\,m_0)\,\vert h_1\vert^2
\nonumber\\
&&-(\alpha_{41}^*\,m_0+\alpha_{60}^*\,\mu_0)\,(h_2.h_1)^*
+\big[\,(\alpha_{50} +2\,\alpha_{60})\,\mu_0+2\alpha_{81}^*\,m_0\big]\,
(h_1.h_2)
\nonumber\\
\rho_{22}
&=&\,\,\,
(2\alpha_{21}^*\,m_0+\alpha_{60}^*\,\mu_0)\vert h_2\vert^2
+(\alpha_{31}^*\,m_0+\alpha_{60}^*\,\mu_0)\,\vert h_1\vert^2
\nonumber\\
&&-\big[(2\alpha_{20}+\alpha_{30})\,\mu_0+\alpha_{61}^*\,m_0\big]\,(h_1.h_2)
+\alpha_{61}^*\,m_0\,(h_2.h_1)^*\qquad\qquad\quad
\label{rhos}
\eea

\medskip\noindent
The non-trivial field-dependent K\"ahler metric gives for the last 
term in $V_F$ of eq.(\ref{VV}):
\medskip
\bea
V_{F,2}&=&
\vert \mu_0\vert^2
\Big[
2\,\big(\alpha_{10}+\alpha_{20}+\alpha_{40}\big)
\vert h_1\vert^2\,\vert h_2\vert^2
+(\alpha_{30}+\alpha_{40})\,\big(\vert h_1\vert^4+\vert h_2\vert^4\big)
\nonumber\\[6pt]
&& +\,2\,\big(\alpha_{10}+\alpha_{20}+\alpha_{30}\big)
\,\vert h_1.h_2\vert^2+\,\,
\big(\vert h_1\vert^2+2\,\vert h_2\vert^2\big)
\big(\alpha_{50}\,h_2.h_1+h.c.\big)
\nonumber\\[6pt]
&&+\,\big(2\vert h_1\vert^2+\vert h_2\vert^2\big)
\big(\alpha_{60}\,h_2.h_1+h.c.\big)\Big]
\label{VF2}
\eea

\medskip\noindent
so that
 $V_F=V_{F,1}+V_{F,2}$. Further, for  the gauge contribution, we have:
\medskip
\bea
V_{gauge}&=&\frac{1}{2}\big( D_w^2+D_Y^2)\,\big[ 
1+ (\alpha_{70}\,h_2.h_1+h.c.)\big]
\nonumber\\[5pt]
&=&
\frac{g_1^2+g_2^2}{8}
\,\big(\vert h_1\vert^2-\vert h_2\vert^2\big)\,
\big[
\big(
1+f_1(h_{1,2}))\,\vert h_1\vert^2-(1+f_2(h_{1,2}))\,
\vert h_2\vert^2\big]
\nonumber\\[3pt]
&+&
\frac{g_2^2}{2}
\,(1+f_3(h_{1,2}))
\vert h_1^\dagger\,h_2\vert^2\qquad\label{VG}
\eea

\medskip\noindent
obtained with (\ref{ddd}) and where $f_{1,2,3}$ are functions of $h_{1,2}$:
\medskip
\bea\label{fs}
f_1(h_{1,2})& \equiv &4\,\alpha_{10}\,\vert h_1\vert^2
+\,\big[\,(2\alpha_{50}-\alpha_{70})\,h_2.h_1+h.c.\big)\big]
\nonumber\\
f_2(h_{1,2})& \equiv &4\,\alpha_{20}\,\vert h_2\vert^2
+\,\big[\,(2\alpha_{60}-\alpha_{70})\,h_2.h_1+h.c.\big)\big]
\nonumber\\
f_3(h_{1,2})& \equiv &\tilde\rho_1+\tilde\rho_2+(\alpha_{70}\,h_2.h_1+h.c.)
\eea
with
\bea\label{tilderhoi}
\tilde\rho_1(h_{1,2})\equiv2\alpha_{10}\,\vert h_1\vert^2 +\alpha_{30}\,\vert
h_2\vert^2+
\big[(\alpha_{50}-\alpha_{70})\,\,h_2.h_1+h.c.\big]
\nonumber\\
\tilde\rho_2(h_{1,2})\equiv2\alpha_{20}\,\vert h_2\vert^2 +\alpha_{30}\,\vert
h_1\vert^2+
\big[(\alpha_{60}-\alpha_{70})\,\,h_2.h_1+h.c.\big]
\eea

The scalar potential also has corrections $V_{SSB}$ from
supersymmetry breaking, due to spurion dependence in
higher dimensional operators (of dimensions $d=5$ and $d=6$);
in addition we also have the usual soft breaking term 
from the MSSM. As a result 
\medskip
\bea
\!\!V_{SSB}\!\!\!&=&- m_0^2\,\Big[
\alpha_{12}\,\,\vert h_1\vert^4
+\,\alpha_{22}\,\,\vert h_2\vert^4
+\,\alpha_{32}\,\,\vert h_1\vert^2\,\vert h_2\vert^2
+\,\alpha_{42}\,\,\vert h_2.h_1\vert^2
\\[3pt]
&&+\,\,\big(\alpha_{52}\,\,\vert h_1\vert^2\,(h_2.h_1)+h.c.\big)
+\big(\alpha_{62}\,\,\vert h_2\vert^2\,(h_2.h_1)+h.c.\big)\Big]
\nonumber\\[3pt]
&&\!\!\!\!-
\Big[\,m_0^2\,\alpha_{82}\,(h_1.h_2)^2+\zeta_{11}\,m_0\,(h_2.h_1)^2
+\mu_0\,B_0\,m_0\,(h_1.h_2)\!+\!h.c.\Big]
+ \! m_0^2\,(c_1 \vert h_1\vert^2+\!c_2 \vert h_2\vert^2)\nonumber
\label{VSSB}
\eea

\medskip\noindent
Finally, in $\cO_{1,...8}$
there are  non-standard kinetic terms that can contribute to
$V$ when the scalar singlet components (denoted $h_i^0$) of $h_i$ 
acquire a vev. The relevant terms are:
\medskip
\bea
\cL_H\supset(\delta_{ij^*}+ g_{ij^*})\,\,
\partial_{\mu}\,h_i^0\,\partial^\mu h_j^{0
  *},\qquad
i,j=1,2.
\eea
where the field dependent metric is:
\bea
g_{11^*}&=& 4\,\alpha_{10}\,\vert h_1^0\vert^2
+(\alpha_{30}+\alpha_{40})\,\vert h_2^0\vert^2
-2\,(\alpha_{50}\, h_1^0\,h_2^0\,+h.c.)
\nonumber\\
g_{12^*}&=&
(\alpha_{30}+\alpha_{40})\,h_1^{0 *}\,h_2^0
-\alpha_{50}^*\,\,h_1^{0 * 2}
-\alpha_{60}\,\,h_2^{0 \, 2},\qquad g_{21^*}=g_{12^*}^*
\nonumber\\
g_{22^*}&=& 4\,\alpha_{20}\,\vert h_2^0\vert^2
+(\alpha_{30}+\alpha_{40})\,\vert h_1^0\vert^2
-2\,(\alpha_{60}\, h_1^0\,h_2^0\,+h.c.)
\eea

\medskip\noindent
For simplicity we only included the $SU(2)$
higgs singlets contribution, that we actually need in the following,
but the discussion can be extended to the general case.
The metric $g_{ij^*}$
is expanded about a background value $\langle h_i^0\rangle
=v_i/\sqrt 2$, then 
 field re-definitions are performed to
obtain canonical kinetic terms; these bring further corrections
to the scalar potential. The field re-definitions are:
\medskip
\bea
h_1^0&\ra& h_1^0\,\,\Big(1-\frac{\tilde g_{11^*}}{2}\Big)\,
-\frac{\tilde g_{21^*}}{2}\,h_2^0
\nonumber\\
h_2^0&\ra& h_2^0\,\,\Big(1-\frac{\tilde g_{22^*}}{2}\Big)\,-
\frac{\tilde g_{12^*}}{2}\,h_1^0,\qquad \tilde g_{ij^*}\equiv 
g_{ij^*}\Big\vert_{h_i^0\rightarrow v_i/\sqrt 2} 
\label{red}
\eea

\medskip\noindent
Since the metric has corrections which are $\cO(1/M^2)$,
after (\ref{red}) only the MSSM soft breaking terms and the
MSSM quartic terms are affected.
The other terms in the scalar potential, already suppressed by
one or more powers of the scale $M$  are affected only beyond
the approximation  $\cO(1/M^2)$ considered here.
Following (\ref{red})
the correction terms $\cO(1/M^2)$  induced by the MSSM quartic terms and
by soft breaking terms in $V_{SSB}$ are:
\medskip
\bea
V_{k.t.}&=& 
\tilde m_1^2\, (- \tilde g_{11}^*) \,\vert  \,h_1^0\,\vert^2
+\tilde m_2^2\,(- \tilde g_{22}^*) \,\vert  \,h_2^0\,\vert^2
-\frac{1}{2}\,\big(\tilde m_1^2+\tilde m_2^2\big)\,
\big(\tilde g_{21^*}\,h_1^{0 *}\,h_2^0 + h.c.\big)
\nonumber\\
&+&
\frac{1}{2}\,
\Big[\,B_0\,m_0\,\mu_0\,\Big(
\,(\tilde g_{11^*}+\tilde g_{22^*})\,\,h_1^0\,h_2^0
+\,\tilde g_{12^*}\,h_1^{0\,2}+\tilde g_{21^*}\, h_2^{0\,2}
\Big)+h.c.\Big]
\nonumber\\
&-& \frac{g^2}{8}\,\big(\,\vert\, h_1^0\,\vert^2-\vert \,h_2^0\,\vert^2\big)
\,\big(\tilde g_{1 1^*}\,\vert \,h_1^0\,\vert^2-\tilde
g_{22^*}\,\vert\,h_2^0\,\vert^2+h.c.\big)
\label{KT}
\eea

\medskip\noindent
Using eqs.(\ref{LL}), (\ref{VV}), (\ref{VF1}), (\ref{VF2}),
(\ref{VG}), (\ref{VSSB}), (\ref{KT}), we find 
the full scalar potential. With the notation
$\tilde m_i^2\equiv c_i m_0^2+\vert \mu_0\vert^2$, 
$i=1,2$  ($c_{1,2}$,  were introduced in $Z_i$ of eq.(\ref{lo}))
one has:
\medskip
\bea\label{vvv}
V&=&V_{F,1}+V_{F,2}+V_{G}+V_{SSB}+V_{k.t.}
\\[5pt]
&=&V_{k.t.}+ \tilde m_1^2 \vert h_1 \vert^2
+\tilde m_2^2 \vert h_2\vert^2
-\big[\mu_0\,B_0\,m_0\,h_1 \cdot h_2
+h.c.\big]\nonumber\\[5pt]
&+&\frac{\lambda_1}{2}\, \,\vert h_1\,\vert^4
+\frac{\lambda_2}{2}\,\,\vert h_2\,\vert^4
+\lambda_3 \,\vert h_1\,\vert^2\,\,\vert h_2\,\vert^2\,
+\lambda_4\,\vert\,h_1\cdot h_2\,\vert^2\nonumber\\[3pt]
&+&
\Big(\,\,
\frac{\lambda_5}{2}\,(h_1\cdot  h_2)^2
+\lambda_6\,\vert\,h_1\,\vert^2\,
(h_1 \cdot h_2)+
\lambda_7\,\vert\,h_2\,\vert^2\,(h_1 \cdot h_2)+h.c.\Big)
\nonumber\\[3pt]
&+&
\frac{g^2}{8}\,\big(\vert h_1\vert^2-\vert h_2\vert^2\big)
\big(f_1(h_{1,2})\,\vert h_1\vert^2-f_2(h_{1,2})\,\vert h_2\vert^2\big)
+4\,\vert \zeta_{10}\vert^2 \vert h_1.h_2\vert^2\,(\vert h_1\vert^2
+\vert h_2\vert^2)
\nonumber\\[3pt]
&+&
\frac{g_2^2}{2}\,f_3(h_{1,2}) \,\vert h_1^\dagger h_2\vert^2
\nonumber
\eea

\medskip\noindent 
where $g^2=g_1^2+g_2^2$, and $f_{1,2,3}(h_{1,2})$
are all quadratic in $h_i$, see eq.(\ref{fs}).
Except $V_{k.t.}$, all other fields are in the SU(2) doublets notation.
The following notation for $\lambda_i$  was introduced:
\medskip
\bea\label{fffV}
{\lambda_1}/{2}&=&
{\lambda_1^0}/{2}
-\vert \mu_0\vert^2\,(\alpha_{30}+\alpha_{40})
-m_0^2\,\alpha_{12}
-2 m_0\,{\rm Re}\big[\alpha_{51}\,\mu_0\big]\\
{\lambda_2}/{2}&=&
{\lambda_2^0}/{2}
-\vert \mu_0\vert^2\,(\alpha_{30}+\alpha_{40})
-m_0^2\,\alpha_{22}
-2 m_0\,{\rm Re}\big[\alpha_{61}\,\mu_0\big]
\nonumber\\
\lambda_3&=&
\lambda_3^0
-\,2\,\,\vert \mu_0\vert^2\,(\alpha_{10}+\alpha_{20}+\alpha_{40})
-m_0^2\,\alpha_{32}
-2 m_0\,{\rm Re}
\big[(\alpha_{51}+\alpha_{61})\,\mu_0\big]
\nonumber\\
\lambda_4&=& \lambda_4^0
-\,2\,\,\vert \mu_0\vert^2\,
(\alpha_{10}+\alpha_{20}+\alpha_{30})
-m_0^2\,\alpha_{42}
-
2\,m_0\,{\rm Re}\big[(\alpha_{51}+\alpha_{61})\,\mu_0\big]
\nonumber\\
{\lambda_5}/{2}&=&-\,m_0\,\mu_0\,(\alpha_{51}+\alpha_{61})-
m_0\,\zeta_{11}-m_0^2\,\alpha_{82}
\nonumber\\
\lambda_6&=&
\,\vert \mu_0\vert^2\,(\alpha_{50}+2\,\alpha_{60})
+m_0^2\,\alpha_{52}
+m_0\,\mu_0\,(2\,\alpha_{11}+\alpha_{31}+\alpha_{41})
+2\,m_0\,\mu_0^*\,\alpha_{81}^*
+\, 2\,\zeta_{10}\,\mu_0^*
\nonumber\\
\lambda_7&=&
\,\vert \mu_0\vert^2\,(\alpha_{60}+2\, \alpha_{50})
+m_0^2\,\alpha_{62}
+m_0\,\mu_0\,(2\,\alpha_{21}+\alpha_{31}+\alpha_{41})
+2\,m_0\,\mu_0^*\,\alpha_{81}^*
+\,2\,\zeta_{10}\,\mu_0^*
\nonumber
\eea

\medskip\noindent
Eq.(\ref{fffV}) shows the effects of various higher dimensional operators
on the scalar potential. As a reminder, note that all $\alpha_{ik}\sim
\cO(1/M^2)$, while $\zeta_{11}, \zeta_{10}\sim \cO(1/M)$. 
The latter can dominate, but this depends on the value of
$\tan\beta$; when this is large, $\cO(1/M^2)$ have comparable size.
In specific models correlations exist among these coefficients.
The above remarks apply to the case when the 
$d=5$ and $d=6$ operators considered
are generated by the same ``new physics'' beyond the MSSM (i.e. are
suppressed by the same scale).
However, as  mentioned earlier, this may not always be the case;
in various models contributions from some $d=6$ operators
can be independent of those from $d=5$ operators (and present even in
the absence of the latter), if generated by different ``new physics''. 
A case by case analysis is then needed for a thorough 
analysis of all possible scenarios for ``new
physics'' beyond the MSSM higgs sector.

We also used the following notation
for the corresponding MSSM contribution:
\medskip\bea\label{ms}
\lambda_1^{0}/2=\frac{1}{8}\,(g_2^2+g_1^{2}),\,\quad
\lambda_2^{0}/2=\frac{1}{8}\,(g_2^2+g_1^{2}),\,\,\quad
\lambda_3^{0}=\frac{1}{4}\,(g_2^2-g_1^{2}),\,\quad
\lambda_4^{0}= -\frac{1}{2}\,g_2^2,\,\quad
\eea

\medskip\noindent
One can include MSSM loop corrections by replacing $\lambda_i^0$ with 
radiatively corrected values \cite{Carena:1995bx}.

The overall sign of the $h^6$  terms depends on the relative size of
$\alpha_{j0}$, $j=1,2,5,6,7$, and cannot be fixed even locally, in the absence
of the exact values of these coefficients of the $d=6$ operators. 
Effective operators of $d=5$, ($\zeta_{10}$),
also contribute to the overall sign, however these alone cannot 
fix it. At large fields values higher and higher dimensional operators
become relevant and contribute to it. We therefore do not impose that
$V$ be bounded from below at large fields. 
For a discussion of stability with $d=5$ operators only
 see \cite{Blum:2009na}.

Eqs.(\ref{vvv}), (\ref{fffV}) of $V$ in the presence of $d=5,6$ effective
operators are  the main result of this section.
For simplicity, one can take $\tilde g_{12^*}$, 
$\tilde g_{21^*}$  real (similar for $B_0\mu_0$), 
possible if for example  $\alpha_{50},\alpha_{60}$
are real and no vev for $Im h_i$; in the next section 
we shall  assume that this is the case.

\section{Corrections to the MSSM Higgs masses: analytical results}
\label{section4}

With the general expression for the scalar
potential we compute the mass spectrum.
From the scalar potential, one evaluates
the mass of  CP-even Higgs fields $h,H$:
\medskip
\bea
m^2_{h,H}\equiv
\frac{1}{2}\frac{\partial^2 V}{\partial h_i^0\partial
  h_j^0}\bigg\vert_{\langle h_i\rangle =v_i/\sqrt 2, 
\langle\,\Im h_i\rangle=0}\eea

\medskip\noindent
In the leading order $\cO(1/M)$ one has (upper signs for $m_h$):
\medskip
\bea\label{mhH}
 m_{h, H}^2\!\!&=&\!\!\!\frac{m_Z^2}{2}
+\frac{B_0\,m_0\mu_0 (u^2+1)}{2\,u} \mp\frac{\sqrt w}{2}
+{  v}^2\,\Big[
\,(2\,\zeta_{10}\,\mu_0)\,\,q_1^{\pm}
+\,(-2\, m_0 \,\zeta_{11})\,\,q_2^{\pm}\Big]+\delta m_{h,H}^2
\qquad\eea

\medskip\noindent
with
\medskip
\bea
q_1^\pm\!\! & = &\!\!\!\frac{1}{4 \, u^2\,(1+u^2)\sqrt w}
\nonumber\\[7pt]
&\times &\!\!\!\!\!
\Big[
- (1-6u^2+u^4)\,u\,\sqrt w\mp 
\Big(m_Z^2 u(1-14 u^2+u^4)-B_0\,m_0\mu_0 (1+u^2) (1+10 u^2
  +u^4)\Big)\Big]
\nonumber\\[5pt]
q_2^\pm\!\! & = &
\mp \frac{2 u}{(1+u^2)^2\sqrt
  w}\,\Big[-B_0\,m_0\mu_0(1+u^2)-m_Z^2\,u\Big]
\quad\eea
where
\bea
w\equiv m_Z^4 + \big[-B_0\,m_0\,\mu_0 (1+u^2)^3+2 m_Z^2 u(1-6 u^2 +u^4)
\big]\frac{(-B_0\,m_0\mu_0)}{u^2(1+u^2)},\qquad u\equiv \tan\beta
\eea
In eq.(\ref{mhH}) 
\bea
\delta m_{h,H}^2= \cO(1/M^2)
\eea
and we also  used that $m_Z=g\,v/2$.
One also shows  that the Goldstone mode has  $m_G=0$ and
the pseudoscalar  A has a mass:
\medskip
\bea\label{ma}
m_A^2=
\frac{1+u^2}{u}\,B_0\,m_0\,\mu_0
-\,\frac{1+u^2}{u}\, \zeta_{10}\,\mu_0\,v^2\,
+2\,m_0\,\zeta_{11}\,v^2+\delta m_A^2,\quad \delta m_A^2=\cO(1/M^2)
\eea

\medskip\noindent
The corrections $\cO(1/M)$ of $m_{h,H}$ and $m_A$ showed
 in (\ref{mhH}), (\ref{ma}),
 agree with earlier findings \cite{Dine}.

Ignoring for the moment the corrections $\cO(1/M^2)$,
one eliminates $B_0$ between (\ref{mhH}) and (\ref{ma})
to obtain:
\medskip
\bea\label{mhold}
 m_{h,H}^2&=&
\frac{1}{2}\Big[m_A^2+m_Z^2\mp\sqrt{\tilde w}\Big]
\nonumber\\[3pt]
&+&
{(2\,\zeta_{10}\,\mu_0)\,
{  v}^2\,\sin 2\beta}
\,\Big[1\pm\frac{m_A^2+m_Z^2}{\sqrt{\tilde w}}\Big]+
\frac{(-2\,\zeta_{11}\,m_0)\,{  v}^2}{2}\,\Big[1\mp
\frac{(m_A^2-m_Z^2)\,\cos^2 2\beta}{\sqrt{\tilde w}}\Big]
\nonumber\\[5pt]
&+&
\delta^\prime m_{h,H}^2, \qquad \qquad  \delta^\prime m_{h,H}^2
= \cO(1/M^2)
\eea

\medskip
\noindent
where the upper (lower) signs correspond to $h$ ($H$) respectively
and
\medskip
\bea
\tilde w\equiv
(m_A^2+m_Z^2)^2-4\,m_A^2\,m_Z^2\,\cos^2 2\beta
\eea

\medskip\noindent
in agreement with \cite{Dine}. This
 is important if one considers $m_A$ as an input; it is also needed
if one considers the limit of large $\tan\beta$ at fixed $m_A$ (see
later).

Regarding the $\cO(1/M^2)$ corrections  of $\delta m_{h,H}^2, \delta
m_A^2$ and $\delta^\prime m_{h,H}^2$ of eqs.(\ref{mhH}),
 (\ref{ma}), (\ref{mhold}) in the general case
of including all operators and their associated supersymmetry breaking,
they have a rather complicated form.
For most purposes, an expansion in $1/\tan\beta$
of $\delta m_{h,H}^2$, $\delta m_A^2$, $\delta^\prime m_{h,H}^2$
 is accurate enough.
The reason for this is that it is only at large $\tan\beta$ that
$d=6$ operators  bring corrections comparable to those of $d=5$
operators. The relative $\tan\beta$ 
enhancement of $\cO(1/M^2)$ operators compensates for the extra 
suppression factor $1/M$ that these operators have relative 
to $\cO(1/M)$ operators (which involve both $h_1$ and $h_2$ and 
thus are not enhanced in this limit). Note however that in some models
 only $d=6$ operators may be present, depending on the details of
the ``new physics'' generating the effective operators.

If we neglect the susy breaking effects of $d=6$ operators (i.e.
$\alpha_{j1}=\alpha_{j2}=0$, $\alpha_{j0}\not=0$, $j=1,...,8$) 
and with $d=5$ operators contribution, one has\footnote{In the
case of including the supersymmetry breaking effects from effective 
operators, associated with coefficients $\alpha_{j1}$, $\alpha_{j2}$
$j=1,2,..8$, the exact formula is very long and is not
included here.} for the correction $\delta m_{h,H}^2$ in eq.(\ref{mhH}) 
(upper signs correspond to $\delta m_h^2$)
\smallskip
\bea\label{dmh}
\delta m_{h,H}^2=\sum_{j=1}^{7}\,\,\gamma^\pm_j\,\,\alpha_{j\,0}
+\gamma^\pm_{x}\,\,\zeta_{10}\,\zeta_{11}  
+\gamma^\pm_{z}\,\,\zeta_{10}^2\,   
+\gamma^\pm_{y}\,\,\zeta_{11}^2     
\label{masshiggs}
\eea

\smallskip\noindent
The expressions of the coefficients $\gamma^\pm$
are provided in Appendix~\ref{appendixC}
and can  be used for numerical studies.
While these expressions are exact, they are complicated 
and not very transparent.
It is then instructive to analyse  an approximation of the 
$\cO(1/M^2)$ correction as an expansion in $1/\tan\beta$. 
We present in this limit the correction $\delta m_{h,H}^2$ 
of eq.(\ref{mhH}), which also includes all  supersymmetry breaking 
effects associated with all $d=5,6$ operators,
(i.e.  $\alpha_{j1}\not=0, \alpha_{j2}\not=0$, 
$\zeta_{11}\not=0$, $j=1,..8$)
in addition to the MSSM soft terms.
This has a simple expression:
\medskip
\bea\label{admh}
\delta m_{h}^2
&=& 
-2 \,v^2\, \Big[
 \alpha_{22} m_0^2+2 \alpha_{61}
\,m_0\mu_0+(\alpha_{30}+\alpha_{40})\,\mu_0^2
-  \alpha_{20} \,m_Z^2\Big]
\nonumber\\[-1pt]
&+&\frac{v^2}{\tan\beta}
\Big[ 4\, \alpha_{62}\,m_0^2+4 \mu_0\,m_0\,( 
2 \alpha_{21}+\alpha_{31}+\alpha_{41}+2 \alpha_{81})
+
4 \mu_0^2\,\,(2 \alpha_{50}+\alpha_{60})
\nonumber\\[-2pt]
&&\qquad\qquad -\,\, m_Z^2\,(2 \alpha_{60}-3 \alpha_{70})
-\frac{v^2}{(B_0m_0\mu_0)}\,(2\zeta_{10}\,\mu_0)^2
\Big]+\cO(1/\tan^2\beta)
\qquad
\eea

\medskip\noindent
which is obtained with $(B_0m_0\mu_0)$ kept fixed. 
The result is dominated by the first line, including both
 susy and non-susy terms from the effective operators.
 This correction can be comparable
to linear terms in $\zeta_{10}$,\,$\zeta_{11}$
from $d=5$ operators for $(2\,\zeta_{10}\mu_0) \approx
1/\tan\beta$  (see later). Not all $\cO_{1,2...8}$
are necessarily present, so in some models some 
$\alpha_{ij}$, $\zeta_{10}$, $\zeta_{11}$
could vanish. 
Also:
\medskip
\bea\label{admH}
\delta m_H^2
&=&-
\frac{1}{4} (B_0m_0\mu_0) \,v^2\,\alpha_{60}\,\tan^2\beta
+\frac{v^2\,\tan\beta}{8}
\Big[
-8 B_0m_0\mu_0\,\alpha_{20} -4 \alpha_{62} m_0^2
\nonumber\\[-2pt]
&-&4 \mu_0\,m_0 ( 2 \alpha_{21}+\alpha_{31}+\alpha_{41}+2 \alpha_{81})
-4 \mu_0^2\,(2 \alpha_{50}+\alpha_{60}) +(2 \alpha_{60}-\alpha_{70})\,m_Z^2\Big]
\nonumber\\[-1pt]
&+&\frac{3}{4}\,B_0m_0\mu_0\,v^2(\alpha_{50}+\alpha_{60})+
\frac{v^2}{8 \tan\beta}\Big[
-8 B_0 m_0\mu_0 \alpha_{10}+ (12\alpha_{52}-16 \alpha_{62}) m_0^2
\nonumber\\[-1pt]
&-& 4 \mu_0 m_0 (-6 \alpha_{11}+8
 \alpha_{21}+\alpha_{31}+\alpha_{41}+2 \alpha_{81})
-4 \mu_0^2(5 \alpha_{50}-2 \alpha_{60})
\nonumber\\[-1pt]
&+&(6 \alpha_{50}+20  \alpha_{60}-13 \alpha_{70})\,m_Z^2
+\frac{8\,v^2}{B_0 m_0 \mu_0}\, (2\,\zeta_{10}\,\mu_0)^2  
\Big]+\cO(1/\tan^2\beta)
\eea

\medskip\noindent
which is  obtained for $(B_0m_0\mu_0)$  fixed. 
Note the $\cO(1/M^2)$ effects from $d=5$ operators ($\zeta_{10}^2$).

Similar expressions exist for the neutral pseudoscalar $A$.
The results are  simpler in this case
and we present the  exact expression
of $\delta m_A^2$ of (\ref{ma}) in the most general case, that  includes
all supersymmetry breaking effects from the 
 operators of $d=5,6$ and from the MSSM. One finds
\medskip
\bea\label{magen}
\delta m_A^2&=&
\frac{v^2}{8 \tan^2\beta\,(1+\tan^2\beta)}\Big[
-\,2 \,B_0\,m_0\mu_0\,\alpha_{50}
+\big[- ( 4 \alpha_{31}+4\alpha_{41}+8 \alpha_{81}+8 \alpha_{11} )
\,m_0\mu_0 
\nonumber\\
&-& 4 \alpha_{52} m_0^2
 - 8B_0 m_0\mu_0 \alpha_{10}- 4\,(\alpha_{50}+2
  \alpha_{60})\mu_0^2 
+(2 \alpha_{50}-\alpha_{70})\,m_Z^2\,\big]\,\tan\beta
\nonumber\\
&+&\big[
2 B_0\,m_0\,\mu_0\,(10 \alpha_{50}+3\alpha_{60})\,
+16 \alpha_{82}m_0^2
+16 (\alpha_{51}+\alpha_{61})m_0\,\mu_0\big]\tan^2\beta
\nonumber\\
&+&\!\!\! 2\,\big[\!
-4 B_0\,m_0\mu_0 (\alpha_{10}+\alpha_{20}+2\,\alpha_{30}+2\,\alpha_{40})
\!-6 (\alpha_{50}+\alpha_{60})\,\mu_0^2
-(\alpha_{50}+\alpha_{60}-\alpha_{70})\,m_Z^2
\nonumber\\
&-& 2 (\alpha_{62}+\alpha_{52})\,m_0^2
-4 (\alpha_{11}+\alpha_{21}+ \alpha_{31}+ \alpha_{41}+2 \alpha_{81})\,
 m_0\mu_0
 \big]\,\tan^3\beta
\nonumber\\
&+&
\big[ 2\,B_0\,m_0\,\mu_0\,(3 \alpha_{50}+10\,\alpha_{60})\,
+16 \alpha_{82} m_0^2+16 (\alpha_{51} +\alpha_{61})\,m_0\mu_0
\big]\tan^4\beta
\nonumber\\
&-&
\big[
8 B_0\,m_0\mu_0\,\alpha_{20}+
4\,(2 \alpha_{50}+\alpha_{60})\,\mu_0^2
-(2 \alpha_{60}-\alpha_{70})\,m_Z^2
+4 \alpha_{62}\,m_0^2
\nonumber\\
&+& 4\,\,(2 \alpha_{21}+\alpha_{31}
+\alpha_{41}+2\alpha_{81})\,m_0\,\mu_0\big]\,\tan^5\beta
-\,\,2 \,B_0\,m_0\,\mu_0\,\alpha_{60}\,\tan^6\beta\,\Big]\qquad
\eea

\medskip\noindent
We also showed that
 $\delta m_G=0$ so the Goldstone mode remains massless
in $\cO(1/M^2)$, which is a good consistency check.
A result similar to that in eq.(\ref{admh}) is found
from an expansion of (\ref{magen}) in the large $\tan\beta$ limit:
\smallskip
\bea
\delta m_A^2&=&
-\frac{1}{4}\,(B_0m_0\mu_0)\,\alpha_{60}\,v^2\,\tan^2\beta
+\frac{\tan\beta}{8}
v^2 \Big[
-8 B_0 m_0\mu_0 \alpha_{20}-4 \alpha_{62} m_0^2\nonumber\\
&-&(8 \alpha_{21} +4 \alpha_{31}+4 \alpha_{41}+8 \alpha_{81})\,m_0\mu_0
-(8 \alpha_{50}+4 \alpha_{60})\mu_0^2+2 \alpha_{60}\,m_Z^2-\alpha_{70}\,m_Z^2
\Big]\nonumber\\
&+&\frac{v^2}{4}
\Big[ B_0m_0\mu_0 (3\alpha_{50} + 11 \alpha_{60}) + 
        8 m_0^2 \alpha_{82} +
 8 m_0\mu_0 (\alpha_{51} + \alpha_{61})\Big]
\nonumber\\
&+&\!\!\!
\frac{v^2}{8\tan\beta}
\Big[- 8B_0m_0\mu_0\,(\alpha_{10}+2 \alpha_{30}+2 \alpha_{40})
 - 4\,(2 \alpha_{11}+ \alpha_{31}+ \alpha_{41}
+ 2\alpha_{81})\,\,m_0\mu_0
\nonumber\\
&-&4 \alpha_{52}\,m_0^2
- ( 4\alpha_{50}+8 \alpha_{60})\mu_0^2
-(2 \alpha_{50}+4 \alpha_{60}-3 \alpha_{70})m_Z^2\Big]
+\cO(1/\tan^2\beta)
\eea

\medskip\noindent
We  emphasise that the large $\tan\beta$ limits presented so far
were done with $(B_0 \,m_0\mu_0)$ fixed. While this is certainly an 
interesting case, because then $m_A$ becomes
large\footnote{and thus likely to re-introduce a little hierarchy to
  explain.}
a more physical case to consider at large $\tan\beta$
 is that in which one keeps $m_A$ fixed ($B_0 m_0\mu_0$ arbitrary). 

We  present below the correction $\cO(1/M^2)$ to $m_{h,H}^2$ for 
the case $m_A$ is kept fixed to an appropriate value. 
The result is (assuming $m_A\!>\!m_Z$, otherwise 
$\delta' m_h^2$ and $\delta' m_H^2$ are exchanged):
\bea\label{dd1}
\delta^\prime m_{h}^2
\!\!\!&=&
-2\, v^2\,\Big[ \alpha_{22} \,m_0^2+(\alpha_{30}+\alpha_{40})\mu_0^2
+2 \alpha_{61} \,m_0\,\mu_0 
- \alpha_{20}\,m_Z^2\Big]
-\frac{(2\,\zeta_{10}\,\mu_0)^2\,\,v^4}{m_A^2-m_Z^2}
\nonumber\\
&+&\!\!\!\!\frac{v^2}{\tan\beta}
\bigg[\frac{1}{(m_A^2-m_Z^2)}
\Big( 4 \,m_A^2\,\big(\,
(2 \alpha_{21}\!+\!\alpha_{31}\!+\!\alpha_{41}\! +\!2 \alpha_{81})
\,m_0\,\mu_0\! +\! (2\alpha_{50}\! +\!\alpha_{60})\,\mu_0^2
+\alpha_{62}\,m_0^2\big)
\nonumber\\
&-&\, (2 \alpha_{60}-3\alpha_{70})\,m_A^2\,m_Z^2
-(2\alpha_{60}+\alpha_{70})\,m_Z^4\Big)
+\frac{8\,(m_A^2+m_Z^2)\,\,(\mu_0\,m_0\,\zeta_{10}\,\zeta_{11})
\,v^2}{(m_A^2-m_Z^2)^2}
\bigg]\nonumber\\
&+&\cO(1/\tan^2\beta)
\eea

\medskip\noindent
A similar formula exists for the correction to $m_H$:
\medskip
\bea\label{dd2}
\delta^\prime m_H^2
\!\!\!&=&
\Big[
-2\, \big( m_0\mu_0\, (\alpha_{51}+\alpha_{61})+\alpha_{82}\,m_0^2
\big)\,v^2 +\frac{(2\,\zeta_{10}\,\mu_0)^2\,v^4}{m_A^2-m_Z^2}
\Big]\nonumber\\
&\!\!\!+&\!\!\!\!\!
\frac{v^2}{\tan\beta}
\!\Big[ 
\frac{1}{m_A^2\!-\!m_Z^2}
\Big(2 m_A^2\,\big(2\, (\alpha_{11}\!-\!\alpha_{21})\,m_0\mu_0
+\!(\alpha_{60}\!-\!\alpha_{50})\,\mu_0^2
+\!(\alpha_{52}\!-\!\alpha_{62})\,m_0^2
-\alpha_{60}\,m_A^2\big)
\nonumber\\
&-&\big[\,
4\, (\alpha_{11}+\alpha_{21}+\alpha_{31}+\alpha_{41}+2
\alpha_{81})\,m_0\mu_0
+
6(\alpha_{50}+\alpha_{60})\,\mu_0^2
+
2(\alpha_{52}+\alpha_{62})\,m_0^2
\nonumber\\[3pt]
&\!\!\!\!-&
\!\!\!(\alpha_{50}\! +\! 5\alpha_{60}\! -\! 2\alpha_{70})\,m_A^2 
\big]\,m_Z^2-(\alpha_{50}\!-\alpha_{60})\,m_Z^4\Big) 
\!-\! \,\frac{8\,(m_A^2+m_Z^2)\,
(\mu_0\,m_0\,\zeta_{10}\,\zeta_{11})
\,v^2}{(m_A^2-m_Z^2)^2}\Big]\nonumber\\
&+&\cO(1/\tan^2\beta)
\eea

\medskip
Corrections (\ref{dd1}), (\ref{dd2})
must be added to the rhs of eq.(\ref{mhold})
to obtain the value  of $m_{h,H}^2$ expressed in function of
$m_A$ fixed. 
The corrections in eqs.(\ref{dmh}) to (\ref{dd2})  
extend the result in \cite{Dine} to include all $\cO(1/M^2)$
terms and represent the main result of this section.

From eqs.(\ref{admh}), (\ref{dd1}) we are able to identify
 the  effective operators of $d=6$ that give the leading contributions  
to $m^2_h$, which is important for model building.
These are $\cO_{2,3,4}$ in the absence of supersymmetry breaking 
and $\cO_{2,6}$ when this is broken, see also eqs.(\ref{operators18}). 
It is however preferable  to increase $m_h^2$ by supersymmetric 
rather than supersymmetry-breaking effects
of the effective operators,  because the latter are less under control
in the effective approach and one would favour a supersymmetric
solution to the  fine-tuning problem associated with increasing
the MSSM Higgs mass above the LEPII bound. Therefore $\cO_{2,3,4}$
are the leading operators, with the remark that $\cO_2$ has a 
smaller effect, of order $(m_Z/\mu_0)^2$ relative to $\cO_{3,4}$
(for similar $\alpha_{j0}$, $j=2,3,4$). At smaller $\tan\beta$,
$\cO_{5,6}$ can also give significant contributions, while $\cO_7$ 
has a relative suppression factor $(m_Z/\mu_0)^2$.

\section{Analysis of the leading corrections and effective operators}
\label{section5}

In general one would expect that $d=6$ operators give sub-leading
contributions to the spectrum, compared to $d=5$ operators,
in the case that all these operators are present and originate from
integrating the same massive ``new physics'' (i.e. are suppressed by
powers of the same scale $M$, which is not always the
case). Even so, for large $\tan\beta$ the latter acquire a relative suppression
factor, and the two classes of operators can indeed give comparable corrections.
At large $\tan\beta$  with $m_A$ fixed,
by comparing $\cO(1/M)$ terms in eq.(\ref{mhold})
against $\cO(1/M^2)$ terms in eqs.(\ref{dd1}), (\ref{dd2}),
one identifies the situation when these two classes of operators
give comparable corrections:
\bigskip
\bea\label{s01}
&&\frac{4 m_A^2}{m_A^2-m_Z^2}\frac{\vert \,\zeta_{10}\,\mu_0
\,\vert}{\tan\beta}
\approx
\bigg\vert
  \alpha_{22} m_0^2+(\alpha_{30}+\alpha_{40})\mu_0^2
+2 \alpha_{61} m_0 \mu_0 
- \alpha_{20}m_Z^2
+\frac{2\,(\zeta_{10}\,\mu_0)^2\, v^2}{m_A^2-m_Z^2}
\bigg\vert\nonumber\\[8pt]
&&\bigg\vert 
\,\,\zeta_{11}\,m_0 \,\,+\frac{4 m_Z^2}{m_A^2-m_Z^2}
\frac{\zeta_{10}\,\mu_0}{\tan\beta}\bigg\vert
\approx
\bigg\vert
\,\big( m_0\mu_0\,(\alpha_{51}+\alpha_{61})+\alpha_{82}\,m_0^2\big)
-\frac{2\,(\zeta_{10}\,\mu_0)^2 \,v^2}{m_A^2-m_Z^2}\bigg\vert\qquad
\eea

\bigskip\noindent
In this case $\cO(1/(M\,\tan\beta))$ 
and $\cO(1/M^2)$ corrections are approximately equal
(for $M\approx m_0\,\tan\beta$).
Similar relations can be obtained from comparing
(\ref{mhH}), (\ref{ma}), against $\delta m_{h,H}^2$ of
(\ref{admh}), (\ref{admH}), (\ref{magen}).
Note that if these relations are satisfied this does not
necessarily mean a failure  of the effective field theory
expansion, since the ``new physics'' that generates these 
operators may be different! Indeed, the corrections
from $d=6$ and $d=5$ operators can be completely
independent (uncorrelated).
However, if all  operators involved in (\ref{s01}) are
generated by the same massive physics,
one would expect that the lhs be smaller than the rhs.
In this case one  obtains 
 conditions for the coefficients of the operators
that should be considered in numerical analyses.
The exact form of such conditions depends on which operators
are present\footnote{As an example, assuming
at least one $d=6$ operator is generated by the same physics as 
the $d=5$ one considered, and if we neglect the supersymmetry breaking
associated effects, then from (\ref{s01})
 $d=6$ operators could give comparable corrections
for $\vert 2\,\zeta_{10}\,\mu_0\,\vert\approx  
g^2/\tan\beta\approx 0.55/\tan\beta$
or $M\approx 2\,\mu_0\,\tan\beta/g^2\approx 3.6\times
\mu_0\,\tan\beta$.}.
Note that  operators with $d>6$ could not acquire a
$\tan\beta$ enhancement relative to $d=6$ operators
to become comparable in size, and they will always have 
an extra suppression factor ($\sim 1/M$).

Let us now examine more closely the corrections to the Higgs masses
due to  $d=6$ operators.  The interest is to maximise the correction
to the MSSM classical value of $m_h$.
From eq.(\ref{admh}) and (\ref{dd1}) and their $\alpha_{ij}$ 
dependence and ignoring susy breaking corrections ($\alpha_{jk}$, $k\not=0$),
we saw that $\cO_{3,4}$  bring the largest 
correction (at large $\tan\beta$), and to a lower extent also 
$\cO_2$. At smaller $\tan\beta$, $\cO_{5,6,7}$ can have significant
corrections. All this can be seen from the relative variation:
\bigskip
\bea\label{opop}
\epsilon_{rel}\!\!&\equiv&
\frac{m_h-m_Z}{m_Z}=\sqrt{\delta_{rel}}-1,\qquad \textrm{where}
\nonumber\\[10pt]
\delta_{rel}
\!\!\!&\equiv&\!\!\!
1-\frac{4 m_A^2}{m_A^2-m_Z^2}\frac{1}{\tan^2\beta}
+\frac{v^2}{m_Z^2}\,\,\bigg\{
\frac{2\,\zeta_{10}\,\mu_0}{\tan\beta}\,\frac{4\,m_A^2}{m_A^2-m_Z^2}
+\frac{(-2\,\zeta_{11}\,m_0)}{\tan^2\beta}\,
\frac{2\,(m_A^4+m_Z^4)}{\,\,(m_A^2-m_Z^2)^2}\,
\nonumber\\[3pt]
&-&\!\Big[2 \,\Big( \alpha_{22}\,m_0^2+(\alpha_{30}+\alpha_{40})\,\mu_0^2
+2\, \alpha_{61}\,m_0\,\mu_0-\alpha_{20}\,m_Z^2\Big)
+\frac{(2\,\zeta_{10}\,\mu_0)^2\,v^2}{m_A^2-m_Z^2}\Big]
\nonumber\\
&+&\frac{1}{\tan\beta}\frac{1}{m_A^2-m_Z^2}
\Big[ 4 \,m_A^2\,\mu_0\,\,
\Big(\,( 2 \alpha_{21}+\alpha_{31}+\alpha_{41}+2\alpha_{81})\,m_0
+(2\,\alpha_{50}+\alpha_{60})\,\mu_0\Big)
\nonumber\\
&+&\!\!\!\! 4\,\alpha_{62}\,m_0^2\,m_A^2
-\! (2 \alpha_{60}\!- \! 3\,\alpha_{70})\,
m_A^2\,m_Z^2-(2\,\alpha_{60}\!+\!\alpha_{70})\,m_Z^4
\!+\! 8 \,\zeta_{10}\,\zeta_{11}\,\mu_0\,m_0\,
v^2\frac{m_A^2\!+\!m_Z^2}{m_A^2\!-\!m_Z^2}\Big]
\bigg\}\nonumber\\[4pt]
&+&
\cO(1/\tan^4\beta)+
\!\cO(\tilde m/(M\tan^3\beta))+\cO(\tilde m^2/(M^2\tan^2\beta))
\eea

\bigskip\noindent
where $\tilde m$ is some generic 
mass scale of the theory such as $\mu_0$, $m_Z$, $m_0$, $v$.
The arguments of the functions $\cO$ in the last line
show explicitly the origin of these corrections (MSSM, $d=5$ or $d=6$
operators, respectively).
Depending on the signs of coefficients $\alpha_{jk}$, $\zeta_{10}$,
$\zeta_{11}$ 
this relative
 variation can be positive and increase $m_h$ above the MSSM 
classical upper bound ($m_Z$). Eq.(\ref{opop})
 gives the overall relative change
of the classical value of $m_h$ in the presence of all possible 
higher dimensional operators of $d=5$ and $d=6$ beyond the MSSM Higgs
sector, for large $\tan\beta$ with $m_A$ fixed. The expansion is
accurate enough to be used also at intermediate 
 $\tan\beta$, but this
also depends on the ratio $\tilde m/M$;
for small $\tan\beta$ the terms in the last line in 
(\ref{opop}) give an error estimate; alternatively one can use
exact $\delta m_{h,H}^2$ in (\ref{dmh}).

A similar result exists for the case the limit of large $\tan\beta$
is taken with $(B_0m_0\mu_0)$ fixed (instead of $m_A$). Then
\medskip
\bea
\delta_{rel}\!\!\! & \equiv &
 1-\frac{4}{\tan^2\beta}
+\frac{v^2}{m_Z^2}
\bigg\{
\frac{4\,(2\,\zeta_{10}\,\mu_0)}{\tan\beta}+
\frac{2}{\tan^2\beta}\,
\Big(\, (-2\,\zeta_{11}\,m_0)   
+\frac{2\,m_Z^2\,(2\,\zeta_{10}\,\mu_0)}{B_0\,m_0\,\mu_0}\Big)
\nonumber\\
&-&2\,\Big[
\alpha_{22}\,m_0^2+2\,\alpha_{61}\,m_0\,\mu_0+(\alpha_{30}+\alpha_{40})
\,\mu_0^2-\alpha_{20}\,m_Z^2\Big]
+\frac{1}{\tan\beta}\Big[\frac{(2\,\zeta_{10}\,\mu_0)^2\,v^2}{-B_0\,m_0\,\mu_0}
\nonumber\\
&+\!\!&4\,( 2\,\alpha_{21}\!+\!\alpha_{31}\!+\!\alpha_{41}
\!+\! 2 \alpha_{81})\,m_0\,\mu_0
+4\,(2\,\alpha_{50}\!+\!\alpha_{60})\,\mu_0^2
\!+\! 4 \alpha_{62}\,m_0^2
-(2\,\alpha_{60}\!-\! 3\alpha_{70})\,m_Z^2
\Big]\bigg\}\nonumber\\[3pt]
&+&
\cO(1/\tan^4\beta)+\cO(\tilde m/(M\tan^3\beta))+
\cO(\tilde
m^2/(M^2\tan^2\beta))\label{pop}
\eea

\medskip\noindent
which can be
 used in numerical applications even for smaller, 
intermediate values of $\tan\beta$.

In (\ref{opop}), (\ref{pop}), the $d=6$ operators ($\alpha_{ij}$
dependence) give contributions
which are dominated by $\tan\beta$-independent terms.
One particular limit to consider  for $\delta m_h^2$ or $\delta'
m_h^2$ is that 
in which the effective operators of $d=6$ have coefficients 
such that these  contributions or those in the first line in (\ref{admh}),
(\ref{dd1})
add up to maximise $\delta_{rel}$. Since coefficients
$\alpha_{ij}$ are not known, as an example we can choose them equal
in absolute value
\bea
-\alpha_{22}=-\alpha_{61}=-\alpha_{30}=-\alpha_{40}=\alpha_{20}>0
\eea
In this case, at large $\tan\beta$:
\bea\label{maxc}
\delta m_h^2\approx 2\, v^2 \alpha_{20}\big[
m_0^2+2\,m_0\mu_0+2\,\mu_0^2+\,m_Z^2\big]
\eea

\medskip\noindent
and similar for $\delta' m_h^2$.
A simple numerical example is illustrative. For $m_0=1$ TeV,
$\mu_0=350$ GeV, and with $v\approx 246$ GeV, one has
$\delta m_h^2\approx 2.36\,\alpha_{20}\,\times 10^{11}$ (GeV)$^2$.
Assuming  $M=10$ TeV and ignoring $d=5$ operators, with 
$\alpha_{20}\sim 1/M^2$ and the MSSM value of $m_h$ taken to be
 its upper classical limit $m_Z$ (reached for large $\tan\beta$),
we obtain an increase of $m_h$ from $d=6$ operators alone 
of about $\Delta m_h=12.15$ GeV to $m_h\approx 103$ GeV.
An increase of $\alpha_{20}$ by a factor of 2.5 to $\alpha_{20}\sim 2.5/M^2$
would give $\Delta m_h\approx 28$ GeV to $m_h\approx 119.2$ GeV,
which is above the  LEPII bound. 
Note that this increase is realised even for a scale $M$ of
``new physics'' beyond the LHC reach. 

Considering instead the larger, loop-corrected MSSM value of $m_h$,
to which we add the $d=6$ operators effects, the relative 
increase  of $\Delta\,m_h$ due to $d=6$ operators alone is mildly 
reduced. However, the effective operators of $d=6$ could in this case reduce the
amount of fine-tuning for the electroweak scale, since these
operators can increase the effective quartic coupling of the Higgs and
thus reduce the fine-tuning, even for a smaller increase of $m_h$.
This was indeed observed for the case of
 $d=5$ operators in the presence of MSSM  quantum corrections to $m_h$,
when the overall mass of $m_h$ (see Appendix~\ref{appendixD}) 
can easily reach values of $130$ GeV
with a reduced, acceptable fine-tuning 
(less than $10$ \cite{r3}) of the electroweak scale \cite{Cassel:2009ps}
(for a scale of effective operators
close to $10$ TeV).  A similar result may be expected in the presence of
$d=6$ operators \cite{cgr}.   Finally, the above choice of $M=10$ TeV was
partly motivated by the fine-tuning  results of \cite{Cassel:2009ps} (valid for
$d=5$ operators) and also on convergence grounds: the expansion
parameter of our effective analysis is $m_q/M$ where $ m_q$ is any scale of the
theory, in particular it can be $m_0$. For a susy breaking scale
$m_0\sim \cO(1)$ TeV (say $m_0=3$ TeV) and $c_{1,2}$ of (\ref{lo})
(or $\alpha_{ij}$ of $\cZ_i(S,S^\dagger)$) 
of order unity (say $c_{1,2}=2.5$) one finds for $M=10$ TeV that
$c_{1,2}\,m_0/M=0.75$ which is already close to unity, i.e. at
the limit of validity of the expansion in powers of $1/M$ of 
the effective approach considered.

From these considerations, one may see that 
effective operators of $d=6$ can indeed bring a significant
increase of $m_h$  to values compatible with the LEPII bound, however, 
the value of the  increase  depends on implicit assumptions, like
the type and number of operators present and 
whether their overall sign as generated by the ``new physics'' is
consistent with an increase of $m_h$. 
Let us briefly refer to this latter issue.

We therefore consider the case of the leading contribution to $m_h$
in the large $\tan\beta$ case. One would prefer to generate,
from a renormalisable model,  the leading operators with 
supersymmetric coefficients satisfying
\bea\label{sign}
\alpha_{20}>0,\, \alpha_{30}<0,\, \alpha_{40}<0
\eea
in order to increase $m_h$.
Let us recall that $\cO_{1,2,3}$ can be easily generated
by integrating out a massive gauge boson $U(1)'$ or  $SU(2)$
triplets \cite{Dine}. $\cO_4$ can be generated by a massive gauge singlet
or $SU(2)$ triplets. Let us 
discuss the signs of the operators when so generated:

\noindent
{\bf (a):} Integrating out a massive vector superfield $U(1)'$
under which Higgs fields have opposite charges (to avoid a
Fayet-Iliopoulos term), 
one finds $\alpha_{20}\!<\!0$ and $\alpha_{30}\!>\!0$ (also $\alpha_{10}\!<\!0$)
\cite{Dine},
which is opposite to  condition  (\ref{sign}). This can 
however be changed, if for example there are additional 
pairs of massive Higgs
doublets also charged under new $U(1)'$, and then $\cO_3$ could be
generated with  $\alpha_{30}<0$.\,\,
{\bf (b):} Integrating massive $SU(2)$ triplets that couple to the MSSM
Higgs sector would bring $\alpha_{20}\!>\!0$, $\alpha_{40}\!<\! 0$,
$\alpha_{30} > 0$, so the first two of these satisfy (\ref{sign}).
{\bf (c):} Integrating a massive gauge singlet would bring
$\alpha_{40}>0$, which would actually 
decrease $m_h$. Finally, at large $\tan\beta$,
due to additional corrections\footnote{Further constraints
exist from $\rho$-parameter: $\rho-1=-v^2/M^2\,
(\alpha_{10}\cos^4\beta+\alpha_{20}\,\sin^4\beta
-\alpha_{30}\,\sin^2\beta\,\cos^2\beta)+\cO(v^4/M^4)$, see
\cite{Brignole:2003cm}, which at large $\tan\beta$ is dominated by 
$\alpha_{20}$, while the effect of  $\alpha_{30}$ is strongly suppressed;
  thus $\alpha_{30}$ is less constrained than $\alpha_{20}$ and 
a better choice for increasing $m_h$.}
that effective operators bring to the $\rho$ parameter
\cite{Brignole:2003cm},
it turns out that $\alpha_{40}$ and $\alpha_{30}$ can have the largest 
correction to $m_h^2$, while avoiding $\rho$-parameter constraints.
The case of a massive gauge singlet or additional $U(1)'$ 
vector superfield (giving $\cO_{3,4}$)
would have the advantage of preserving 
gauge couplings unification at one-loop.

For smaller $\tan\beta$, operators $\cO_{5,6,7}$ could bring
significant corrections to $m_h$; it is more difficult to generate
these in a renormalisable set-up, when more additional states are needed.
For example $\cO_{5,6}$ can be generated by integrating out a pair
of massive Higgs doublets and a massive gauge singlet, but 
the overall sign of $\alpha_{50,60}$ would 
depend on the details of the model. This discussion shows that
while effective operators can in principle increase $m_h$,
deriving a detailed, renormalisable model where they are generated 
with appropriate signs for their (supersymmetric) coefficients
is not a simple issue. These examples are however rather naive and 
other generating mechanisms for $\cO_i$ could be in place 
(in a renormalisable set-up\footnote{For some models with extended 
MSSM Higgs sector see
\cite{Espinosa:1998re,Espinosa:1991gr,Espinosa:1992sk,Espinosa:1992hp}.}) 
with  appropriate signs to increase $m_h$.

\section{Conclusions}

We investigated in detail the Higgs sector of the MSSM in the
presence of all $d=5$ and $d=6$ effective operators that can be
present in this sector. This was motivated by the attempt to 
better understand the MSSM Higgs sector and its consistency with
the quantum stability of the electroweak scale, the
associated amount of fine tuning,  
the LEPII bound on $m_h$ and the so far negative searches for 
TeV scale  supersymmetry. New physics beyond the current 
MSSM Higgs sector, parametrised by these effective operators,
could  alleviate  these problems while retaining
at the same time the advantages of low-energy supersymmetry,
which was a main motivation of this work. The effective 
operators description used here is little dependent on the exact
 details of the new physics which generates  these operators.

Two classes of such effective operators were present and investigated:
higher dimensional derivative and non-derivative operators.
We showed in Appendix~\ref{appendixB}
 that the former can be removed from the action through
appropriate non-linear field redefinitions and this is essentially
equivalent to integrating out the massive additional ghost 
degrees of freedom (of mass $\sim M$) that such operators bring.
It was also clarified in Appendix~\ref{appendixB} that the use of ``on-shell'' 
setting of these operators brings similar results 
and is  appropriate only in the leading order in the suppressing scale.
The remaining, non-derivative operators contribute to the Higgs sector
and their effects on the scalar potential and on the CP even and odd
Higgs masses were computed analytically.

Despite their suppression by an extra
power of the high scale $M$ relative to the $d\!\!~=~\!\!5$ operators, the
relative $\tan\beta$ enhancement of the $d=6$ operators
compensates for this suppression, to bring corrections comparable 
to those of the  $d=5$ operators, in the case both classes of
operators are generated from the same high energy physics.
This may not always be the case and it 
is possible that some of the $d=6$ operators be present even
in the absence of the $d=5$ operators, if these classes of operators
 are generated
by different new physics beyond the MSSM Higgs sector. Since our
analysis assumed independent coefficients for all operators (whether 
of $d=5$ and $d=6$), our results are general and can be applied even if only
some of these operators are present, regardless of their origin.

We identified the effective
operators which give the most significant contributions to $m_h$
in the limit of large $\tan\beta$ and these can be both
supersymmetric and non-supersymmetric. The supersymmetric
case  is preferable and also more important since such contribution would
essentially alleviate a problem of fine-tuning which is intrinsically
susy-breaking related. Of these operators $\cO_{3,4}$ would have 
the advantage of avoiding further $\rho$-parameter constraints.
At small $\tan\beta$ other operators ($\cO_{5}$, $\cO_{6}$,
$\cO_{7}$) could bring relevant corrections to $m_h$.
Numerically, the impact of $d=6$ operators alone on the mass of the
lightest Higgs can be  in the region of  $10-30$ GeV. 
In the presence of MSSM loop effects and 
eventually $d=5$ operators (if also present), this effect can help
 keep a low electroweak scale fine tuning,
while respecting the LEPII mass bound and the current bounds on
superpartners masses. If a  larger increase of $m_h$ is 
sought from ``new physics'' beyond MSSM, the effective approach may not be reliable,
and one should instead consider other approaches, 
such as MSSM with additional
light states which are {\it not} integrated out.

Simple possibilities were listed for the ``new physics'' that, upon 
being integrated out, could generate  these operators, 
in a renormalisable set-up.
The ``new physics'' could be associated with the presence of a
massive  gauge singlet ($\cO_4$), massive $U(1)'$ ($\cO_{1,2,3}$),
massive  $SU(2)$ triplets ($\cO_{1,2,3,4}$). Some of these 
cases can have difficulties, through their impact on unification, 
perturbativity up to the Planck scale, etc. Of these,
a very interesting possibility is that of extra $U(1)'$ massive 
gauge boson or massive gauge singlet,  which do not share these 
difficulties in the leading order. 
In the simplest mechanisms 
generating the corresponding, leading operators $\cO_{3,4}$, 
the overall coefficients of their
supersymmetric part have however signs  opposite to those 
needed to maximise the classical correction to the lightest Higgs mass 
(at large $\tan\beta$). Nevertheless such operators could
be generated  in other ways, when 
correlations among the coefficients of
the effective operators could also be present.
The next step in this analysis would 
be to construct a renormalisable model that would generate
in the  effective action such  operators
with appropriate values for their coefficients.

\bigskip\bigskip
\bigskip\bigskip
\bigskip\noindent
{\bf Note added:}\newline
 While this paper was being typewritten, a similar study appeared
\cite{Carena:2009gx} which has a partial overlap with this work.

\bigskip\bigskip\bigskip
\section*{Acknowledgements}

This work was partially supported by ANR (CNRS-USAR) 
contract 05-BLAN-007901,   INTAS grant 03-51-6346,
contract PITN-GA-2009-237920,  MRTN-CT-2006-035863, CNRS PICS no. 3747
and 4172 and the ERC Advanced Grant - 226371 (``MassTeV'').\,
E.D. thanks the GGI Institute in Florence and the Aspen Center for
Physics for hospitality during the completion of this work.  
P.T. thanks the ``Propondis'' Foundation for the financial support
during the last stages of this work.\,
D.G. thanks the CERN Theory Group and \'Ecole Polytechnique  Paris 
for the financial support and S.~Cassel,
 C.~Grojean and  G.G.~Ross  for interesting related
discussions.


\section{Appendix:}

\def\theequation{A-\arabic{equation}}
\def\thesubsection{A}
\setcounter{equation}{0}

\subsection{Integrals of operators $\cO_{1,..8}$:}
\label{appendixA}

\begin{eqnarray} 
\mathcal{O}_{1} &=&
\frac{1}{M^{2}}\int d^{4}\theta \,\,
\mathcal{Z}_{1}(S,S^{\dagger })\,\,
(H_{1}^{\dagger }\,e^{V_{1}}\,H_{1})^{2} 
\nonumber \\
&=& 
2 \alpha_{1 0}\,
\Big[
(h_1^\dagger h_1)\,\big[\,(\cD_\mu h_1)^\dagger\,(\cD^\mu h_1)
+h_1^\dagger\,\frac{D_1}{2}\,h_1 + F_1^\dagger F_1\,\big]+
\vert\,h_1^\dagger F_1\vert^2 
+
( h_1^\dagger  \cD^\mu h_1)
(h_1^\dagger \overleftarrow{\cD_\mu} h_1)
\Big]
\nonumber\\
&+&
\Big[2\,\alpha_{1 1}\,m_0 \,(h_1^\dagger h_1)(F_1^\dagger h_1)+h.c.\Big]
+\alpha_{12}\, m_0^2\,(h_1^\dagger h_1)^2+{\rm fermionic \,\,part}
\\[3pt]
\mathcal{O}_{2} 
&=&\frac{1}{M^{2}}\int d^{4}\theta \,\,
\mathcal{Z}_{2}(S,S^{\dagger })\,\,
(H_{2}^{\dagger }\,e^{V_{2}}\,H_{2})^{2} 
\nonumber \\
&=& 
2 \alpha_{2 0}\,\Big[(h_2^\dagger h_2)\,\big[\,
(\cD_\mu h_2)^\dagger\,(\cD^\mu h_2)
+h_2^\dagger\,\frac{D_2}{2}\,h_2 + F_2^\dagger F_2\,\big]+
\vert  h_2^\dagger F_2\vert^2 
+
(h_2^\dagger  \cD^\mu h_2)
(h_2^\dagger \overleftarrow{\cD_\mu} h_2)
\Big]
\nonumber\\
&+&
\Big[2\,\alpha_{2 1}\,m_0\, 
(h_2^\dagger h_2)(F_2^\dagger h_2)+h.c.\Big]
+\alpha_{22}\,m_0^2\, (h_2^\dagger h_2)^2+{\rm fermionic \,\,part}
\\[3pt]
\mathcal{O}_{3} &=&
\frac{1}{M^{2}}\int d^{4}\theta \,\,\mathcal{Z} _{3}(S,S^{\dagger
})\,\,
(H_{1}^{\dagger 
}\,e^{V_{1}}\,H_{1})\,(H_{2}^{\dagger }\,e^{V_{2}}\,H_{2}),
\nonumber\\
&=&
\alpha_{30}\,\Big\{
 (h_1^\dagger h_1)\,
\Big[(\cD_\mu h_2)^\dagger\,(\cD^\mu h_2)
+h_2^\dagger\,\frac{D_2}{2}\,h_2 + F_2^\dagger F_2\Big]+
(h_1^\dagger  F_1)(F_2^\dagger h_2)+(1\leftrightarrow 2)\Big\}
\nonumber\\
&+&
\alpha_{30}\,\Big[
(h_1^\dagger \cD_\mu
h_1)(h_2^\dagger\overleftarrow \cD^\mu h_2)
+h.c.\Big]+\Big\{\alpha_{31}\,m_0\, \Big[
(h_1^\dagger h_1)(F_2^\dagger h_2)
+(h_2^\dagger h_2)(F_1^\dagger h_1)
\Big]+h.c.\Big\}
\nonumber\\
&+&\alpha_{32}\,m_0^2\,
(h_1^\dagger h_1)(h_2^\dagger h_2)+{\rm fermionic \,\,part}
\\[3pt]
\mathcal{O}_{4} &=&\frac{1}{M^{2}}\int d^{4}\theta \,\,
\mathcal{Z}_{4}(S,S^{\dagger })
\,\,(H_{2}\,.\,H_{1})\,(H_{2}\,.\,H_{1})^{\dagger },
\nonumber\\
&=&
\alpha_{40}\,\,\partial_\mu (h_2.h_1)\,\partial^\mu(h_2.h_1)^\dagger
+\Big[\alpha_{41}\,m_0\,(h_2.h_1)\,(h_2.F_1 +F_2.h_1)^\dagger +h.c.\Big]
\nonumber\\
&+&\alpha_{42}\,m_0^2\,(h_2.h_1)\,(h_2.h_1)^\dagger
+\alpha_{40}\,\vert h_2\cdot F_1+F_2\cdot h_1\vert^2
+{\rm fermionic \,\,part}
\\[3pt]
\mathcal{O}_{5} &=&\frac{1}{M^{2}}\int d^{4}\theta \,\,\mathcal{Z}%
_{5}(S,S^{\dagger })\,\,(H_{1}^{\dagger 
}\,e^{V_{1}}\,H_{1})\,H_{2}.\,H_{1}+h.c.  \nonumber \\
&=&
\alpha_{50}
\Big\{\Big[
(\cD_\mu h_1)^\dagger\,(\cD^\mu h_1)
+h_1^\dagger\,\frac{D_1}{2}\,h_1 + F_1^\dagger F_1\Big](h_2.h_1)
+
(h_1^\dagger \overleftarrow\cD_\mu h_1)\,\partial^\mu(h_2.h_1)\Big\}
\nonumber\\
&+&\!\!\!
\Big[\alpha_{50} \,(F_1^\dagger h_1)
+\alpha_{51}^*\,m_0\,(h_1^\dagger\,h_1)\Big]\,(h_2.F_1+F_2.h_1)
+
m_0\,\Big[\alpha_{51}\,(F_1^\dagger h_1)+\alpha_{51}^*\,
(h_1^\dagger F_1)\Big]\,(h_2.h_1)
\nonumber\\
&+&
\alpha_{52}\,m_0^2\,(h_1^\dagger h_1)\,(h_2.h_1)+{\rm h.c.\,of \,all}
+{\rm fermionic \,\,part}
\\[3pt]
\mathcal{O}_{6} &=&
\frac{1}{M^{2}}\int d^{4}\theta \,\,\mathcal{Z}_{6}(S,S^{\dagger
})\,\,
(H_{2}^{\dagger }\,e^{V_{2}}\,H_{2})\,\,H_{2}.\,H_{1}+h.c.  \nonumber \\
&=&
\alpha_{60}
\Big\{\Big[
(\cD_\mu h_2)^\dagger\,(\cD^\mu h_2)
+h_2^\dagger\,\frac{D_2}{2}\,h_2 + F_2^\dagger F_2\Big](h_2.h_1)
+
(h_2^\dagger \overleftarrow\cD_\mu h_2)\,\partial^\mu(h_2.h_1)\Big\}
\nonumber\\
&+&\!\!\!
\Big[\alpha_{60} \,(F_2^\dagger h_2)
+\alpha_{61}^*\,m_0\,(h_2^\dagger\,h_2)\Big]\,(h_2.F_1+F_2.h_1)
+
m_0\,\Big[\alpha_{61}\,(F_2^\dagger h_2)+\alpha_{61}^*\,
(h_2^\dagger F_2)\Big]\,(h_2.h_1)
\nonumber\\
&+&
\alpha_{62}\,m_0^2\,(h_2^\dagger h_2)\,(h_2.h_1)+{\rm h.c.\,of \,all}
+{\rm fermionic \,\,part}
\eea
\bea
\mathcal{O}_{7} &=&\frac{1}{M^{2}}\frac{1}{16 g^2 \kappa}
\int d^{2}\theta \,\,
\mathcal{Z}_{7}(S,0)\,\,{\rm Tr}\,\, W^{\alpha }\,W_{\alpha }\,(H_{2}\,H_{1})+h.c. 
\nonumber\\[4pt]
&=&
\frac{1}{2}\,(D_w^2+D_Y^2)\,
\Big[\alpha_{70}\,(h_2.h_1)+\alpha_{70}^*\,(h_2.h_1)^\dagger\Big]
+{\rm fermionic \,\,part}\\[3pt]
\mathcal{O}_{8} &=&
\frac{1}{M^{2}}\int d^{4}\theta
 \,\,\Big[\mathcal{Z}_{8}(S,S^{\dagger })\,\,
\,[(H_{2}\,H_{1})^{2}+h.c.]\Big] 
\nonumber\\[4pt]
&=&
2\,\alpha_{81}^* \,m_0\,(h_2.h_1)\,(h_2.F_1+F_2.h_1)
+m_0^2\,\alpha_{82}\,(h_2\cdot h_1)^2+h.c.+
{\rm fermionic \,\,part}\qquad
\end{eqnarray}%

\medskip
\noindent
 $W^\alpha$ is the susy field strength
of $SU(2)_L$ ($U(1)_Y$) vector superfield $V_w$ ($V_y$) of 
auxiliary  component $D_w$ ($D_Y$). Also
\bea
(1/M^2)\,\,\cZ_i(S,S^\dagger)=\alpha_{i0}
+\alpha_{i1}\,m_0\,\theta\theta
+\alpha_{i1}^*\,m_0\,\overline\theta\overline\theta
+\alpha_{i2}\,m_0^2\,\theta\theta\overline\theta\overline\theta
\eea
and
$\cD^\mu h_i=(\partial^\mu+i/2\,V^\mu_i)\,h_i$,\,
$h_i^\dagger \overleftarrow\cD^\mu=(\cD^\mu h_i)^\dagger$.
Further,  $D_1\equiv \vec D_w\,\vec T+(-1/2)\,\,D_Y$ and
  $D_2\equiv \vec D_w\,\vec T+(1/2)\,\,D_Y$, $T^a=\sigma^a/2$.
Finally, one rescales in all $\cO_i$ ($i\not=7$):\,\,\,
$V_{w}\ra 2\,g_2\,V_w$, 
$V_{y}\ra 2\,g_1\,V_y$.
Then $V_{1,2}= 2\,g_2\,\vec V_w\,\vec T +2\,g_1\,(\mp 1/2)\,V_y$ with the
upper sign (minus) for $V_1$, where
 $V_{1,2}$ enter the definition of $\cO_{1,2}$.
Other notations used above:
$H_1.H_2=\epsilon^{ij}\,H_1^i\,H_2^j$. Also
$\vert h_1\cdot h_2\vert^2
=\vert h_1^i\,\epsilon^{ij}\,h_2^j\vert^2
=\vert h_1\vert^2\,\vert h_2\vert^2
-\vert h_1^\dagger \,h_2\vert^2;$
 $\epsilon^{ij}\,\epsilon^{kj}=\delta^{ik}$;\,\,
$\epsilon^{ij}\,\epsilon^{kl}=
\delta^{ik}\,\delta^{jl}-\delta^{il}\,\delta^{jk}$, $\epsilon^{12}=1$,
with
\bea
h_1=\left(\begin{array}{c}
h_1^0 \\[-1pt]
h_1^-\\
\end{array}\right)
\equiv 
\left(\begin{array}{c}
h_1^1 \\[-1pt]
h_1^2\\
\end{array}\right),\,\,Y_{h_1}=-1;
\qquad
h_2=\left(\begin{array}{c}
h_2^+\\[-1pt]
h_2^0
\end{array}\right)
\equiv
\left(\begin{array}{c}
h_2^1\\[-1pt]
h_2^2
\end{array}\right),\,\,\,Y_{h_2}=+1
\eea

\medskip\noindent
Lagrangian (\ref{LL}) with the above $\cO_{1,..,8}$  leads to
 \medskip
 \bea\label{arhos}
 F_1^{* q}&=&
 -\big\{\epsilon^{qp}\,h_2^p\, \big[
 \mu_0+2\,\zeta_{10}\,(h_1.h_2)
 +\rho_{11}\big]+h_1^{* q}\,\rho_{12}\big\}
 \nonumber\\
 F_2^{* q}&=&
 -\big\{\epsilon^{pq}\,h_1^p \,\big[
 \mu_0+2\,\zeta_{10}\,(h_1.h_2)
 +\rho_{21}\big]+h_2^{* q}\,\rho_{22}\big\}
 \eea

\medskip\noindent
where $\rho_{ij}$ are functions of $h_{1,2}$, given in 
eq.(\ref{rhos0}), (\ref{rhos}). Similarly
\medskip
 \bea\label{ddd}
 D_w^a&=&-g_2\,\Big[\,\,h_1^\dagger T^a\,h_1\,\,(1+\tilde
 \rho_1)+h_2^\dagger\,T^a\,h_2\,\,(1+\tilde \rho_2)\,\Big],
 \qquad T^a=\sigma^a/2
 \nonumber\\
D_Y&=&-g_1\,\Big[\,\,h_1^\dagger \frac{-1}{2}\,h_1\,\,(1+\tilde
\rho_1)+h_2^\dagger\,\frac{1}{2}\,h_2\,\,(1+\tilde \rho_2)\,\Big]
\eea

\medskip\noindent
with notation (\ref{tilderhoi}). This gives
\medskip
\bea
 D_w^a\,D_w^a&=&
 \frac{g_2^2}{4}\,\big[\,\,\big(
 (1+\tilde\rho_1)\,\vert h_1\vert^2-
 (1+\tilde\rho_2)\,\vert h_2\vert^2\big)^2
 +4\,(1+\tilde\rho_1)(1+\tilde\rho_2)\,\vert h_1^\dagger\,h_2\vert^2
\big]
\nonumber\\
D_Y^2&=&
\frac{g_1^2}{4}\,\big(
(1+\tilde\rho_1)\,\vert h_1\vert^2-
(1+\tilde\rho_2)\,\vert h_2\vert^2\big)^2\label{dsq}
\eea

\medskip\noindent
used in the text, eq.(\ref{VG}).

\newpage

\bigskip
\def\theequation{B-\arabic{equation}}
\def\thesubsection{B}
\setcounter{equation}{0}

\subsection{Integrating out the ghosts, field redefinitions,
and ``on-shell'' operators.}\label{appendixB}
\medskip

Here it is shown that operators of $d=5$ or $d=6$ of type 
$\cO_{9,..,15}$ 
encountered in (\ref{der0}) or similar,
which contain higher derivatives, can  be
``removed'' from the action: (1) by integrating out the
ghost degrees of freedom,
(2): using the eqs of motion to set ``on-shell'' the derivative
operator, or (3) by using non-linear field re-definitions.
Beyond the leading order  method (2) is not always applicable, as
showed later for  $d=5$ effective operators (Appendix~\ref{appendixB2})
 and thus it should be used with care.

\subsubsection{The case of $d=6$ operators.}\label{appendixB1}

Let us consider first the case of $d=6$ operators. We use
here method (1) and (2). Similar results are found with method (3).

\medskip\noindent
{\it (1) Integrating out the (super)ghosts.}

\noindent
Take\footnote{A very similar treatment follows if one considers in 
(\ref{pp1}) an opposite sign in front of $\Box/M^2$ \cite{ADG}.}
\bea\label{pp1}
\cL=\int d^4\theta
\Big[\,\Phi^\dagger\,(1+\Box/M^2)\,\Phi+S^\dagger S\Big]
+\bigg\{\int d^2\theta\,\,W[\Phi,S]+h.c.\bigg\}+\cO(1/M^3)
\eea
where $\Box\equiv -1/16\,\overline D^2 \,D^2$ and $S$ denotes in this
appendix  some arbitrary superfield. 
The derivative operator is similar to $\cO_9$ in the absence of
gauge interactions; here we show how to remove this operator.
$W$ can contain non-renormalisable terms up to $\cO(1/M^3)$.
This $\cL$ can be re-written as a second order theory
(for details see \cite{ADG}) with a Lagrangian:
\bea\label{tt1}
\cL&=&\int d^4\theta\,\Big[\Phi_1^\dagger \Phi_1
-\Phi_2^\dagger\Phi_2-\Phi^\dagger_3\,\Phi_3+S^\dagger S\Big]
\nonumber\\
&+&
\int d^2\theta \,\,\Big[\mu_{13}\,\Phi_1\,\Phi_3
+\mu_{23}\,\Phi_2\,\Phi_3+W[\Phi(\Phi_{1,2,3});S]\Big]
+h.c.+\cO(1/M^3)\qquad
\eea
where
\bea
\Phi(\Phi_{1,2,3})=\eta^{-1/4}\,(\Phi_2-\Phi_1),
\qquad
\eta\equiv 
1+4\,m^2/M^2
\eea
and
\bea
\mu_{13}=\mu_{31}=\frac{1-\sqrt\eta}{2\,\eta^{1/4}} \,M=
-\frac{m^2}{M}+\cO(1/M^3)
\nonumber\\
\mu_{23}=\mu_{32}=-\frac{1+\sqrt\eta}{2\,\eta^{1/4}}\,M=
-M+\cO(1/M^3)
\eea

\medskip\noindent
We can integrate out the massive super-ghosts by using their eqs of motion:
\medskip
\bea
&&\frac{1}{4}\,\overline D^2\Phi_2^\dagger
+\mu_{23}\,\Phi_3+\eta^{-1/4}\,W'=0
\nonumber\\
&&\frac{1}{4}\,\overline D^2
\Phi_3^\dagger+\mu_{13}\,\Phi_1+\mu_{23}\,\Phi_2
=0
\eea
giving
\bea
\Phi_3&=& \frac{1}{M}\,W'[-\Phi_1;S]+\cO(1/M^3)\nonumber\\
\Phi_2&=& \frac{1}{4\,M^2}\,\overline D^2\,W^{' \dagger}[-\Phi_1;S]
-\frac{m^2}{M^2}\,\Phi_1+\cO(1/M^3)
\eea

\medskip\noindent
where the derivatives are taken wrt the first argument.
We have
\medskip
\bea
\mu_{23}\,\Phi_2\,\Phi_3&=&-\frac{1}{M^2}\,\bigg[\frac{1}{4}\,\overline
  D^2\,W^{'
    \dagger}[-\Phi_1;S]-m^2\,\Phi_1\bigg]\,W'[-\Phi_1;S]+\cO(1/M^3)
\nonumber\\
\mu_{13}\,\Phi_1\,\Phi_3&=& -\frac{m^2}{M}\,\Phi_1\,W'[-\Phi_1;S]+
\cO(1/M^3)
\nonumber\\
W[\Phi(\Phi_{1,2,3});S]&=&W[-\Phi_1;S]
+\frac{1}{4\,M^2}\,W'[-\Phi_1;S]\,\overline D^2\,W^{'
  \dagger}[-\Phi_1;S]
+\cO(1/M^3)
\eea

\medskip\noindent
Using these one finds
\medskip
\bea\label{ee1}
\cL&=&\int d^4\theta \,\bigg[
\Phi_1^\dagger\Phi_1-\frac{1}{M^2}\,\,W^{' \dagger}[-\Phi_1;S]\,\,
W^{'}[-\Phi_1;S]+S^\dagger S\bigg]
\nonumber\\[7pt]
&+&
\bigg\{\int d^2\theta\,\,\,W[-\Phi_1;S]+h.c.\bigg\}+\cO(1/M^3)\qquad
\eea

\medskip\noindent
This result is valid at energy scales well below the mass of the
ghost $M$.
A similar result is obtained {\it in this leading order},
by using the equations of motion  to set on-shell the higher 
derivative term,  (see below).

\bigskip\noindent
{\it (2) Using eqs of motion to set ``on-shell'' the operators.}

\medskip\noindent
 Consider again (\ref{pp1})
\medskip
\bea
\cL=\int d^4\theta
\Big[\Phi^\dagger\,\Phi -1/(16\,M^2)\,
\overline D^2 \Phi^\dagger\,D^2\Phi + S^\dagger S\Big]
+\int d^2\theta\,\,W[\Phi,S]+h.c.+\cO(1/M^3)
\eea

\medskip\noindent
which can be re-written by using the eqs of motion for $\Phi$
\medskip
\bea
\overline D^2 \Phi^\dagger=4 \,W'[\Phi;S]+\cO(1/M^2)
\eea
to find\medskip
\bea
\cL &=&
\int d^4\theta\,\, \Big[\Phi^\dagger\Phi-\frac{1}{M^2}\,\,
W^{' \dagger}[\Phi;S]\,\,W^{'}[\Phi;S]+S^\dagger S\Big]
\nonumber\\[7pt]
&+&\int d^2\theta \,\,\,W[\Phi;S]+h.c.
+\cO(1/M^3)
\eea

\medskip\noindent
where under the derivative of the superpotential one should include
only the renormalisable terms of $W$, which is correct under the
approximation considered. This result is in agreement with that
of (\ref{ee1}), up to a trivial field redefinition.

\vspace{0.7cm}

\subsubsection{The case of $d=5$ operators.}\label{appendixB2}

We extend the previous discussion to 
the case of extra derivatives in the superpotential
and we take the lowest order case ($d=5$).
Start with
\medskip
\bea\label{ss1}
\cL&=&\int d^4\theta \Big[ \Phi^\dagger\Phi+S^\dagger S\Big]+
\bigg\{\int d^2\theta \Big[ \frac{\sigma}{M}\,\Phi\,\Box\,\Phi
+W[\Phi;S]\Big]+h.c.\bigg\}
\nonumber\\[7pt]
&=& 
\int d^4\theta \Big[ \Phi^\dagger \Phi+
\frac{\sigma}{4M}\big(\Phi D^2\Phi+h.c.\big)\Big]
+\bigg\{\int d^2\theta\,\, W[\Phi;S]+h.c.\bigg\}
\eea

\medskip\noindent
with $\sigma=\pm 1$ and where it is assumed that the superpotential part
of the action can contain additional higher dimensional 
(non-derivative) operators which have mass dimensions $d\leq 5$.
It is shown that one can remove these operators via field
redefinitions
or via integrating out the ghost degree of freedom. These methods 
are shown to be equivalent. In the {\it leading order only}
 setting ``on-shell'' the operator via the eqs of motion also gives
a similar, correct result.

\bigskip\medskip\noindent
{\it (1). Integrating out the (super)ghosts:  }
\medskip
\bea
\cL&=&\int d^4\theta
\,\Big[\Phi_1^\dagger\Phi_1-\Phi_2^\dagger\Phi_2+S^\dagger S\Big]
\nonumber\\[7pt]
&+&\bigg\{\int d^2\theta\,\, \bigg[
\,\,\frac{1}{2}
\,d_2\,\Phi_2^2+d_3\,\Phi_1\,\Phi_2+\frac{1}{2}\,d_1\,\Phi_1^2
+W[\Phi(\Phi_{1,2});S]\,\bigg]+h.c.\bigg\}
\eea
with 
\bea
d_1& =& \frac{(\sqrt{\eta'}-1)^2}{8 \,\sigma \, \sqrt{\eta'}}\,M
=\cO(1/M^3),
\nonumber\\[2pt]
d_2&= &  \frac{(\sqrt{\eta'}+1)^2}{8 \,\sigma \, \sqrt{\eta'}}\,M
=\sigma\,M/2+\cO(1/M^3),\nonumber\\[2pt]
d_3&=&\frac{\eta'-1}{8\,\sigma\,\sqrt{\eta'}}\,\,M\,\,=
(k\,\sigma)\,\,m^2/M+\cO(1/M^3)
\eea
where  $k=17/32$ and
\bea
\Phi\equiv
\eta^{' -1/4}\,(\Phi_2-\Phi_1),\qquad\eta'\equiv 1+(17/4)\,m^2/M^2,\qquad
\eea
We can integrate out the massive ghost superfield $M\gg m$ using
\medskip
\bea
&&\frac{1}{4} 
\overline
D^2\Phi_2^\dagger+\frac{\sigma}{2}\,M\,\Phi_2+W'[\Phi;S]\,\eta^{' -1/4}
+\frac{\sigma\,k\,m^2}{M}\,\,\Phi_1+\cO(1/M^3)=0
\eea

\medskip\noindent
Denote in the following $W'\equiv W'[-\Phi_1;S]$ 
where the derivative is wrt the first argument. Then
\bea
\Phi_2&=&
- \frac{2\,\sigma}{M}\,W'+\frac{4}{M^2}\,W'\,W''\,
- \frac{2\,k\,m^2}{M^2}\,\Phi_1+
\frac{1}{M^2}\,\overline D^2 \,W^{' \dagger}+\cO(1/M^3)
\eea

\medskip\noindent
Taylor expand:
\medskip
\bea
\Phi=-\Phi_1-\frac{2\,\sigma}{M}\,W'+\frac{4}{M^2}\,W'\,W''
+\frac{1}{M^2}\,\overline D^2\,W^{' \dagger}
+\cO(1/M^3)
\eea

\medskip
\noindent
then Taylor expand  $W[\Phi;S]$ in function of $W'$, $W''$ to $\cO(1/M^3)$;
also 
\medskip
\bea
\Phi_2^\dagger\Phi_2&=&\frac{4}{M^2}\,W'\,W^{' \dagger}+\cO(1/M^3)
\nonumber\\[4pt]
d_3\,\Phi_1\Phi_2&=&-\frac{2 \,k\,m^2}{M^2}\,\Phi_1\,W'+\cO(1/M^3)
\nonumber\\[4pt]
\frac{1}{2}\,d_2\,\Phi_2^2&=&
\frac{\sigma}{M}\,W^{' 2}-\frac{4}{M^2}\,W^{' 2}\,W''
+\frac{2 \,k\,m^2}{M^2}\,\Phi_1\,W'-\frac{1}{M^2}\,W'\,\overline D^2\,W^{'
  \dagger}
+\cO(1/M^3)\nonumber\\[4pt]
W[\Phi;S]&=&
W-\frac{2\,\sigma}{M}\,W^{' 2}
+\frac{6}{M^2}\,W^{' 2}\,W''
+\frac{1}{M^2}\,W'\,\overline D^2\,W^{' \dagger}
+
\cO(1/M^3)
\eea

\medskip\noindent
Add everything together
\medskip
\bea\label{lastres}
\cL&=&\int d^4\theta \,\Big[\Phi_1^\dagger\Phi_1
-\frac{4}{M^2}\,W'\,W^{' \dagger}
+S^\dagger \,S\Big]
\nonumber\\[7pt]
&+&
\bigg\{\int d^2\theta \,
\bigg[
W
-\frac{\sigma}{M}\,W^{' 2}+\frac{2}{M^2}\,W^{''}\,W^{' 2}
\bigg]+h.c.\bigg\}+\cO(1/M^3)
\eea

\medskip\noindent
where the argument of  $W$, $W'$, $W''$ above is $[-\Phi_1;S]$
and derivatives are wrt the first argument.
This is equivalent to the starting Lagrangian, with new (non-renormalisable)
interactions but no derivative ones.

\medskip\noindent
{\it (2) Removing derivative operators using field redefinitions.}

\medskip\noindent
Let us show that a similar result is obtained if we use general, 
local field redefinitions.
Start again with:
\medskip
\bea\label{ss1p}
\cL&=&\int d^4\theta \Big[ \Phi^\dagger\Phi+S^\dagger S\Big]+
\bigg\{\int d^2\theta \Big[ \frac{\sigma}{M}\,\Phi\,\Box\,\Phi
+W[\Phi;S]\Big]+h.c.\bigg\}
\nonumber\\[3pt]
&=& 
\int d^4\theta \Big[ \Phi^\dagger \Phi+
\frac{\sigma}{4M}\big(\Phi D^2\Phi+h.c.\big)\Big]
+\bigg\{\int d^2\theta\,\, W[\Phi;S]+h.c.\bigg\}
\eea

\medskip\noindent
First eliminate the $\cO(1/M)$ terms by redefinition:
\bea
\Phi\ra \Phi-\frac{\sigma}{4\,M}\,\overline D^2\Phi^\dagger
\eea
The Lagrangian becomes:
\medskip
\bea
\cL&=&
\int d^4\theta \Big[ \Phi^\dagger \Phi+S^\dagger\,S
+\frac{\sigma}{M}\,\big(W'\,\Phi^\dagger+h.c.\big)
-\frac{3}{32\,M^2}\big(\Phi\,D^2\overline
D^2\,\Phi^\dagger+h.c.\big)\nonumber\\[3pt]
&-&
\frac{1}{8\,M^2}\big(W''\,\Phi^\dagger
\overline D^2\Phi^\dagger+h.c.\big)\Big]
+\bigg\{\int d^2\theta\,\, W[\Phi;S]+h.c.\bigg\}
\eea
Next
\bea
\Phi\ra \Phi-\frac{\sigma}{M}\,W'[\Phi;S]
\eea
which gives
\medskip
\bea
\cL&=&\!\!\!
\int d^4\theta \Big[ \Phi^\dagger \Phi+S^\dagger\,S
-\!\frac{1}{M^2}\,\big(\Phi^\dagger\,W'\,W''+h.c.\big)
-\!\frac{1}{M^2}\,W'\,W^{' \dagger}
-\!
\frac{3}{32\,M^2}\big(\Phi\,D^2\overline
D^2\,\Phi^\dagger+h.c.\big)\nonumber\\[3pt]
&-&
\frac{1}{8\,M^2}\big(W''\,\Phi^\dagger\overline D^2\Phi^\dagger+h.c.\big)
\Big]
+\bigg\{\int d^2\theta\,\, \bigg[W-\frac{\sigma}{M}\,W^{'
  2}+\frac{1}{2\,M^2}\,W^{' 2}\,W''\bigg]
+h.c.\bigg\}\quad
\eea

\medskip\noindent
We eliminate now the $1/M^2$ terms by
\medskip
\bea
\Phi\ra \Phi+\frac{3}{32}\frac{1}{M^2}\,\overline D^2\,D^2\,\Phi
\eea
giving
\bea
\cL&=&
\int d^4\theta \Big[ \Phi^\dagger \Phi+S^\dagger\,S
-\frac{1}{M^2}\,\big(\Phi^\dagger\,W'\,W''+h.c.\big)
-\frac{1}{M^2}\,W'\,W^{' \dagger}-\frac{3}{8\,M^2}\,(W'\,D^2\Phi+h.c.)
\nonumber\\[3pt]
&-&
\frac{1}{8\,M^2}\big(W''\,\Phi^\dagger\overline D^2\Phi^\dagger+h.c.\big)
\Big]
+\bigg\{\int d^2\theta\,\, \bigg[W-\frac{\sigma}{M}\,W^{'
  2}+\frac{1}{2\,M^2}\,W^{' 2}\,W''\bigg]
+h.c.\bigg\}
\eea

\medskip\noindent
Next consider
\bea
\Phi\ra \Phi+\frac{1}{8\,M^2}\,W''\,\overline D^2\Phi^\dagger
\eea

\medskip\noindent
to find

\bea
\cL&=&\!\!\!
\int d^4\theta \Big[ \Phi^\dagger \Phi+S^\dagger\,S
-\frac{3}{2\,M^2}\,\big(\Phi^\dagger\,W'\,W''+h.c.\big)
-\frac{1}{M^2}\,W'\,W^{' \dagger}-\frac{3}{8\,M^2}\,(W'\,D^2\Phi+h.c.)
\Big]\nonumber\\[7pt]
&+&
\bigg\{\int d^2\theta\,\, \bigg[W-\frac{\sigma}{M}\,W^{'
  2}+\frac{1}{2\,M^2}\,W^{' 2}\,W''\bigg]
+h.c.\bigg\}
\eea
then

\bea
\Phi\ra \Phi+\frac{3}{8\,M^2}\,\overline D^2\,W^{' \dagger}[\Phi;S]
\eea
to obtain

\bea
\cL&=&
\int d^4\theta\, \Big[ \Phi^\dagger \Phi+S^\dagger\,S
-\frac{3}{2\,M^2}\,\big(\Phi^\dagger\,W'\,W''+h.c.\big)
-\frac{1}{M^2}\,W'\,W^{' \dagger}\Big]
\nonumber\\[7pt]
&+&
\bigg\{\int d^2\theta\,\, \bigg[W-\frac{\sigma}{M}\,W^{'
  2}+\frac{1}{2\,M^2}\,W^{' 2}\,W''+\frac{3}{8\,M^2}\,W'\,\overline
  D^2\,W^{' \dagger}
\bigg]
+h.c.\bigg\}
\eea

\medskip\noindent
Finally
\medskip
\bea
\Phi\ra \Phi+\frac{3}{2\,M^2}\,W'[\Phi;S]\,W''[\Phi;S]
\eea
one finds
\medskip
\bea\label{second}
\cL&=&\int d^4\theta\, \Big[ \Phi^\dagger \Phi+S^\dagger\,S
-\frac{4}{M^2}\,W'\,W^{' \dagger}\Big]
\nonumber\\[7pt]
&&\qquad\qquad+\,\,
\bigg\{\int d^2\theta\,\, \bigg[W-\frac{\sigma}{M}\,W^{'
  2}+\frac{2}{M^2}\,W^{' 2}\,W''\bigg]
+h.c.\bigg\}\qquad
\eea

\medskip\noindent
which agrees with the result in (\ref{lastres}) up to
and including $\cO(1/M^2)$ terms.

\newpage

\medskip\noindent
{\it (3) Setting ``on-shell'' the derivative 
operators by using the eqs of motion.}

\medskip\noindent
Let us discuss what happens if in action (\ref{ss1}) we
``removed'' the higher derivative terms by 
 using the equations of motion i.e. setting them ``on-shell''. 
It will turn out that only in
leading order ($1/M$) do we obtain a similar $\cL$ as in previous cases.
The eq of motion is
\bea
\overline D^2\Phi^\dagger =4 \,W'+\cO(1/M)
\eea
where $W'\equiv \partial W[\Phi;S]/\partial \Phi$.
This is used in (\ref{ss1}), and after an additional shift to re-write
higher dimensional D-terms as F-terms
\bea
\Phi\ra \Phi-(\sigma/M)\,W'
\eea
we obtain:
\medskip
\bea\label{onsh}
\cL&=&\int d^4\theta
\Big[\Phi^\dagger\Phi+S^\dagger S\Big]+
\bigg\{\int d^2\theta \,\,W[\Phi-(\sigma/M)\,\,W';\,S]+h.c.\bigg\}
\nonumber\\[7pt]
&=&\int d^4\theta
\Big[\Phi^\dagger\Phi+S^\dagger S\Big]+
\bigg\{\int d^2\theta\,\,\Big[ \,W[\Phi;S]-(\sigma/M)
\,\,W'^2[\Phi;S]\Big]+h.c.\bigg\}
\eea

\medskip\noindent
where we used a Taylor expansion in the last step.
If we include the next order, after using
\bea
\overline D^2\Phi^\dagger =4 \,W'-\frac{\sigma}{M}\,\overline D^2\,W^{'
  \dagger}+\cO(1/M^2)
\eea
and after the following redefinitions
\bea
\Phi\ra \Phi-\frac{\sigma}{M}\,W'
\eea
and 
\bea
\Phi\ra \Phi\!+\!\frac{1}{M^2}\,W'\,W''
\eea
one finds
\bea
\cL\!=\!\int \!\!d^4\theta\, \Big[ \Phi^\dagger \Phi+S^\dagger\,S
-\frac{3}{M^2}\,W'\,W^{' \dagger}\Big]
\!+\!
\bigg\{\int d^2\theta \bigg[W\!-\!\frac{\sigma}{M}\,W^{'
  2}\!+\!\frac{3}{2\,M^2}\,W^{' 2}\,W''\bigg]
+h.c.\bigg\}\,\,
\eea

\medskip\noindent
Although this agrees with (\ref{lastres}) in $\cO(1/M)$, it
disagrees with it in order $\cO(1/M^2)$. The reason for this is
that the ``on-shell'' setting method of higher dimensional operators 
is {\it derived} using general field redefinitions only 
in the leading order $\cO(1/M)$. As a result this method should be
used with care.

\def\theequation{C-\arabic{equation}}
\def\thesubsection{C}
\setcounter{equation}{0}
\subsection{Coefficients for the Higgs masses.}\label{appendixC}

The coefficients in eq.(\ref{dmh}) have the following expressions:
\bea
\gamma_1^\pm&=&
\frac{  \pm v^2}{2 u^2 (1+u^2)^3\,w^{ 1/2}}
\nonumber\\
&\times &
\,\Big[(B_0m_0\mu_0)^2 \,(1+u^2)^4- 2 m_Z^2\,u^2 \,
\big[m_Z^2 (1-u^2)^2+(1+u^2)\,( 8 \mu_0^2\,u^2 \pm (u^2-1)\,w^{ 1/2}))
\big]\nonumber\\
&+&(B_0\,m_0\mu_0)\,u (1+u^2)^2\big[m_Z^2 \,(1+u^2)-(\pm\, w^{1/2} (1+u^2) +
16\mu_0^2\,u^2 )\big]
\Big]
\\
\gamma_2^\pm&=&
\frac{\pm v^2}{2 (1+u^2)^3 \,w^{ 1/2}}
\nonumber\\
&\times &
\Big[
(B_0m_0\mu_0)^2 (1+u^2)^4-2 m_Z^2 u^2 \big[8 \mu_0^2 (1+u^2)
+ m_Z^2 (1-u^2)^2 \pm w^{ 1/2}(1-u^4) \big]
\nonumber\\
&-&(B_0 m_0\mu_0)\,u\,(1+u^2)^2\big[
16 \mu_0^2 - m_Z^2 (1+u^2) \pm(1+u^2)\,w^{ 1/2}\big]\Big]
\\
\gamma_3^\pm\!&=&\!\!\gamma_4^\pm=\!
\frac{\pm v^2}{u\,(1+u^2)^2\,w^{1/2}}
\,\big\{\mu_0^2
\big[\!-\!B_0m_0\mu_0\,(1\!+\!u^2)^3\! +m_Z^2 u(1-\!6 u^2\! +u^4)\mp 
u(1\!+\!u^2)^2\,w^{1/2}\big]\nonumber\\
&+& B_0 m_0\mu_0 \,u^2\,(1+u^2)\,m_Z^2
+m_Z^2\,u^3\,(m_Z^2\mp w^{ 1/2})\big\}
\\[7pt]
\gamma_5^\pm&=&
\frac{\mp v^2}{8 u^3\,(1+u^2)^3\,w^{ 1/2}}
\Big[
 (B_0 m_0 \mu_0)^2 (1+u^2)^4\,(-1 +3 u^2)
- (B_0m_0\mu_0)\,u (1+u^2)^2\nonumber\\
&\times &\!\!\! \big[- 2 m_Z^2 (1+5 u^2)
+2\mu_0^2 \, (1+8 u^2 +25 u^4+2 u^6)
 \pm (1+u^2) (3 u^2-1)\,w^{ 1/2}\big]
\nonumber\\
&- &u^2\,m_Z^2 \big[
m_Z^2 (1-19 u^2 -u^4 +3 u^6)
- 2 \mu_0^2\,(1+u^2)(1 -16\,u^2-23 u^4+2 u^6)
\nonumber\\
&\pm &(1+u^2)^2(1+3 u^2)\,w^{ 1/2}\big]
+ 2 \mu_0^2\, u^2\,\big[\pm (1+u^2)^2\,(1-9 u^2 +2 u^4)
w^{ 1/2}\,\big]
\,\,\Big]
\\[7pt]
\gamma_6^\pm&=&
\frac{\pm v^2}{8 u^2\,(1+u^2)^3\,w^{ 1/2}}
\Big[
 (B_0 m_0 \mu_0)^2 \,u\,(1+u^2)^4\,(-3 + u^2)
- (B_0m_0\mu_0)\, (1+u^2)^2\nonumber\\
&\times &\!\!\! \big[ 2 m_Z^2 (5+ u^2)\,u^4
-2\mu_0^2 \, (2+25 u^2 +8 u^4+ u^6)
 \pm (1+u^2)\, (u^2-3)\,u^2\,w^{ 1/2}\big]
\nonumber\\
&+ &u\,m_Z^2 \big[
m_Z^2 (3- u^2 -19 u^4 + u^6)\,u^2
-2 \mu_0^2\,(1+u^2)(2 -23\,u^2-16 u^4+ u^6)
\nonumber\\
&\pm &u^2\,(1+u^2)^2(3+ u^2)\,w^{ 1/2}\big]
- 2 \mu_0^2\, u\,\big[\pm (1+u^2)^2\,(2-9 u^2 + u^4)
w^{ 1/2}\,\big]
\,\,\Big]\\[7pt]
\gamma_7^\pm&=&
\frac{\mp v^2 m_Z^2}{16 u^2 (1+u^2)^3\,w^{ 1/2}}
\Big[
-B_0 m_0\mu_0\,(1+u^2)(1+40 u^2 -114 u^4 +40 u^6+u^8)
\nonumber\\
&+& m_Z^2 \,(u+30 u^5 +u^9) \pm
u (1+u^2)^2 (1-10 u^2 +u^4) \,w^{ 1/2}
\Big]
\\[7pt]
\gamma_{x}^\pm&=&\!\!\!\frac{\pm 8\,(u^2-1)^2\,v^4}{u\,(1+u^2)^3\,w^{ 3/2}}
\, \,\big[ m_Z^2\,u-B_0m_0\mu_0\,(1+u^2)\big]
\big[ 2\,m_Z^2\,u-B_0m_0\mu_0\,(1+u^2)\big]\,m_0\,\mu_0
\\[7pt]
\gamma_{y}^\pm&=& \mp\frac{(-1+u^2)^2\,v^4}{(1+u^2)^4\,w^{ 3/2}}
\,\,\big[m_Z^2\,u-B_0\,m_0\,\mu_0\,(1+u^2)\big]^2\,(4 \,m_0^2)
\eea\bea
\gamma_{z}^\pm&=&\frac{ \mp v^4}{\mu_0^2\,u^2\,(1+u^2)^3\,w^{ 3/2}}
\\
&\times&\!\!\Big[
- 2\,(B_0m_0\mu_0)^3 \,u\,(1+u^2)^4
+m_Z^4\,u^2(1+u^2)
\big(4\,\mu_0^2(-1+u^2)^2-u^2 (2 m_Z^2\pm w^{1/2})\big)
\nonumber\\
&+&\!\! 2\, B_0 m_0\mu_0\,
m_Z^2\,u\,\big[-2 \mu_0^2(u^4-1)^2
+u^2 (m_Z^2(1-14 u^2+u^4)  \pm (u^4 - 6 u^2+1)\,w^{ 1/2}
)\big]\nonumber\\
&+&\!\!
(B_0 m_0\mu_0)^2\,(1+u^2)\big[\mu_0^2
\,(u^4-1)^2+u^2 (2m_Z^2\,(1-14 u^2+u^4) \mp(1+u^2)^2\,w^{1/2})\big]
\Big](4\mu_0^2)\nonumber
\eea

\bigskip
\def\theequation{D-\arabic{equation}}
\def\thesubsection{D}
\setcounter{equation}{0}

\subsection{One-loop $m_h$ with $d=5$ operators}\label{appendixD}
\medskip

For future reference, it is worth mentioning the value of
$m_h$ in the presence of one-loop corrections from top-stop 
and  $d=5$ operators \cite{Cassel:2009ps},
mentioned in the text:
\medskip
\begin{eqnarray} \label{mh1loop5}
m_{h}^{2}\!\!\! &=&\!\!\!\frac{1}{2}\Big[m_{A}^{'\, 2}+m_{Z}^{2}
-\sqrt{{\tilde w}^{'}}+\xi \Big]  
\nonumber  \label{finalmh} \\
&\!\!\!+&\!\!\!\!\!
{(2\,\zeta_{10}\,\mu_0) {\ v}^{2}\,\sin 2\beta }\,
\bigg[1+\frac{m_{A}^{'\, 2}+m_{Z}^{2}}{
\sqrt{\tilde w^{'}}}\bigg]+\frac{(-2\,\zeta_{11}\,m_0)\,{\ v}^{2}}{2}\,\bigg[1-
\frac{(m_{A}^{'\,2}-m_{Z}^{2})\,\cos ^{2}2\beta }{\sqrt{\tilde w^{'}}}\bigg] 
\qquad
\end{eqnarray} 
where 
\begin{eqnarray} 
{\tilde w}^{'}
 &\equiv &[(m_{A}^{'\,2}-m_{Z}^{2})\,\cos 2\beta +\xi ]^{2}+\sin ^{2}2\beta 
\,(m_{A}^{'\,2}+m_{Z}^{2})^{2}  \nonumber  \label{mhf} \\[5pt]
m_{A}^{'\,2} &=&\tilde{m}_{1}^{2}+\tilde{m}_{2}^{2}+\xi /2+
(2\,\zeta_{10}\mu_0)\,v^{2}\,\sin 2\beta +\zeta _{11}\,m_0\,v^{2};\quad 
\xi \equiv \delta\, m_{Z}^{2}\,\sin ^{2}\beta
\end{eqnarray}

\medskip\noindent
where $\delta$ is the one-loop correction from top-stop Yukawa
sector to $\lambda_2^0$ of (\ref{ms}) which changes according 
to $\lambda_2^0\ra
\lambda_2^0\,(1+\delta)$ where \cite{Giudice:2006sn,Carena:1995bx}
\medskip  
\begin{eqnarray} 
\delta  &=&\frac{3\,h_{t}^{4}}{g^{2}\,\pi ^{2}\,}\bigg[\ln \frac{M_{\tilde{t}
}}{m_{t}}+\frac{X_{t}}{4}+\frac{1}{32\pi ^{2}}\,\Big(3\,h_{t}^{2}-16
\,g_{3}^{2}\Big)\Big(X_{t}+2\ln \frac{M_{\tilde{t}}}{m_{t}}\Big)\ln \frac{M_{
\tilde{t}}}{m_{t}}\bigg],  \nonumber\\[6pt]
X_{t} &\equiv &\frac{2\,(A_{t}\,m_{0}-\mu \cot \beta )^{2}}{M_{\tilde{t}}^{2}
}\,\,\Big[1-\frac{(A_{t}\,m_{0}-\mu \cot \beta )^{2}}{12\,\,M_{\tilde{t}}^{2} 
}\,\Big]. 
\end{eqnarray}

\medskip\noindent 
with $M_{\tilde{t}}^{2}\equiv m_{\tilde{t}_{1}}\,m_{\tilde{t}_{2}}$, 
and $g_{3}$ the QCD coupling. 
The combined effect of $d=5$ operators and top Yukawa coupling $h_t$
is that $m_h$ can reach values of $130$ GeV for $\tan\beta\leq 7$ 
with a small fine-tuning $\Delta\leq 10$ \cite{Cassel:2009ps}
and with the supersymmetric coefficient 
$\zeta_{10}$ giving a larger effect than the non-susy one, 
$\zeta_{11}$. Even for a modest increase of $m_h$ from $d=5$ operators
alone  of order $\cO(10 GeV)$, their impact on the effective
quartic coupling of the Higgs field is  significant (due to the small value of 
the MSSM gauge couplings), and this explains the reduction
of fine-tuning by the effective operators.


\begin{thebibliography}{99}

\bibitem{higgsboundLEP} 
R.~Barate \textit{et al.} [LEP Working Group for 
Higgs boson searches], ``Search for the standard 
model Higgs boson at LEP,'' Phys.\ Lett.\ B \textbf{565}, 61 (2003)
 [arXiv:hep-ex/0306033]; S.~Schael  \textit{et al.}
 [ALEPH Collaboration], ``Search for neutral MSSM Higgs bosons at
 LEP,'' Eur.\ Phys.\ J.\ C \textbf{47}, 547 (2006) 
[arXiv:hep-ex/0602042]. 


\bibitem{AbdusSalam:2009qd}
  S.~S.~AbdusSalam, B.~C.~Allanach, F.~Quevedo, F.~Feroz and M.~Hobson,
  ``Fitting the Phenomenological MSSM,''
  arXiv:0904.2548 [hep-ph].
 
\bibitem{Barbieri:1998uv} 
R.~Barbieri and A.~Strumia, ``About the 
fine-tuning price of LEP,'' Phys.\ Lett.\ B \textbf{433} (1998) 63 
[arXiv:hep-ph/9801353]. 
 

\bibitem{Chankowski:1998xv}
 P.~H.~Chankowski, J.~R.~Ellis, M.~Olechowski and 
S.~Pokorski, ``Haggling over the fine-tuning price of LEP,''
 Nucl.\ Phys.\ B  
\textbf{544} (1999) 39 [arXiv:hep-ph/9808275].  
 
\bibitem{Chankowski:1997zh}
 P.~H.~Chankowski, J.~R.~Ellis and S.~Pokorski, 
``The fine-tuning price of LEP,'' Phys.\ Lett.\ B \textbf{423} (1998) 327 
[arXiv:hep-ph/9712234]. 
 

\bibitem{Kane:1998im}
 G.~L.~Kane and S.~F.~King, ``Naturalness implications 
of LEP results,'' Phys.\ Lett.\ B \textbf{451} (1999) 113 
[arXiv:hep-ph/9810374]. 
 
\bibitem{Giudice:2006sn}
 G.~F.~Giudice and R.~Rattazzi, ``Living dangerously 
with low-energy supersymmetry,'' Nucl.\ Phys.\ B \textbf{757} (2006) 19 
[arXiv:hep-ph/0606105]. 
 
\bibitem{Giudice:2007fh}
  G.~F.~Giudice, C.~Grojean, A.~Pomarol and R.~Rattazzi,
  ``The Strongly-Interacting Light Higgs,''
  JHEP {\bf 0706} (2007) 045
  [arXiv:hep-ph/0703164].

\bibitem{Dine} M.~Dine, N.~Seiberg and
 S.~Thomas, ``Higgs Physics as a 
Window Beyond the MSSM (BMSSM),'' 
Phys.\ Rev.\ D \textbf{76} (2007) 095004 
[arXiv:0707.0005 [hep-ph]]. 
 
\bibitem{ADGT}
  I.~Antoniadis, E.~Dudas, D.~M.~Ghilencea and P.~Tziveloglou,
  ``MSSM with Dimension-five Operators (MSSM$_5$),''
  Nucl.\ Phys.\  B {\bf 808} (2009) 155
  [arXiv:0806.3778 [hep-ph]].

\bibitem{Antoniadis:2008ur}
  I.~Antoniadis, E.~Dudas, D.~M.~Ghilencea and P.~Tziveloglou,
  ``Higher Dimensional Operators in the MSSM,''
  AIP Conf.\ Proc.\  {\bf 1078} (2009) 175
  [arXiv:0809.4598 [hep-ph]].

\bibitem{Blum:2009na}
  K.~Blum, C.~Delaunay and Y.~Hochberg,
  ``Vacuum (Meta)Stability Beyond the MSSM,''
  arXiv:0905.1701 [hep-ph].

\bibitem{Berg:2009mq}
  M.~Berg, J.~Edsjo, P.~Gondolo, E.~Lundstrom and S.~Sjors,
  ``Neutralino Dark Matter in BMSSM Effective Theory,''
  arXiv:0906.0583 [hep-ph].
 
 \bibitem{Blum:2008ym} K.~Blum and Y.~Nir, ``Beyond MSSM Baryogenesis,'' 
 Phys.\ Rev.\ D \textbf{78} (2008) 035005 [arXiv:0805.0097 [hep-ph]].  

\bibitem{Ham:2009gu}
  S.~W.~Ham, S.~a.~Shim and S.~K.~OH,
  ``Possibility of spontaneous CP violation 
in Higgs physics beyond the minimal
  supersymmetric standard model,''
  arXiv:0907.3300 [hep-ph].
 
\bibitem{Cassel:2009ps}
  S.~Cassel, D.~M.~Ghilencea and G.~G.~Ross,
  {\it ``Fine tuning as an indication of physics beyond the MSSM,''}
  Nucl.\ Phys.\  B {\bf 825} (2010) 203
  [arXiv:0903.1115 [hep-ph]].

\bibitem{Batra:2003nj} 
P.~Batra, A.~Delgado, D.~E.~Kaplan and T.~M.~P.~Tait, 
``The Higgs mass bound in gauge extensions of the minimal supersymmetric 
standard model,'' JHEP \textbf{0402} (2004) 043 [arXiv:hep-ph/0309149].  

\bibitem{r3} 
R.~Barbieri and G.~F.~Giudice, ``Upper Bounds On Supersymmetric Particle 
Masses,'' Nucl.\ Phys.\ B \textbf{306} (1988) 63;  
 

\bibitem{Georgi:1991ch}
  H.~Georgi,
  ``On-Shell Effective Field Theory,''
  Nucl.\ Phys.\  B {\bf 361} (1991) 339.

\bibitem{Politzer:1980me}
  H.~D.~Politzer,
  ``Power Corrections At Short Distances,''
  Nucl.\ Phys.\  B {\bf 172} (1980) 349.

\bibitem{Arzt}
  C.~Arzt,
  ``Reduced effective Lagrangians,''
  Phys.\ Lett.\  B {\bf 342} (1995) 189
  [arXiv:hep-ph/9304230].

\bibitem{ADG}
  I.~Antoniadis, E.~Dudas and D.~M.~Ghilencea,
  ``Supersymmetric Models with Higher Dimensional Operators,''
  JHEP {\bf 0803} (2008) 045
  [arXiv:0708.0383 [hep-th]].


\bibitem{Piriz:1997id}
  D.~Piriz and J.~Wudka,
  ``Effective operators in supersymmetry,''
  Phys.\ Rev.\  D {\bf 56} (1997) 4170
  [arXiv:hep-ph/9707314].

\bibitem{Carena:1995bx} 
M.~S.~Carena, J.~R.~Espinosa, M.~Quiros and 
C.~E.~M.~Wagner, ``Analytical expressions for radiatively corrected Higgs 
masses and couplings in the MSSM,'' Phys.\ Lett.\ B \textbf{355} (1995) 209 
[arXiv:hep-ph/9504316]. 


\bibitem{cgr}
S.~Cassel, D.~M.~Ghilencea, G.~G.~Ross, work in progress.

\bibitem{Brignole:2003cm}
  A.~Brignole, J.~A.~Casas, J.~R.~Espinosa and I.~Navarro,
  ``Low-scale supersymmetry breaking: Effective description, electroweak
  breaking and phenomenology,''
  Nucl.\ Phys.\  B {\bf 666} (2003) 105
  [arXiv:hep-ph/0301121].
%

\bibitem{Espinosa:1998re}
  J.~R.~Espinosa and M.~Quiros,
  ``On Higgs Boson Masses In Nonminimal Supersymmetric Standard Models,''
  Phys.\ Lett.\  B {\bf 279} (1992) 92.

\bibitem{Espinosa:1991gr}
  J.~R.~Espinosa and M.~Quiros,
  ``Gauge unification and the supersymmetric light Higgs mass,''
  Phys.\ Rev.\ Lett.\  {\bf 81} (1998) 516
  [arXiv:hep-ph/9804235].

\bibitem{Espinosa:1992sk}
  J.~R.~Espinosa and M.~Quiros,
  ``Higgs boson bounds in nonminimal supersymmetric standard models,''
  arXiv:hep-ph/9208226.

\bibitem{Espinosa:1992hp}
  J.~R.~Espinosa and M.~Quiros,
  ``Upper bounds on the lightest Higgs boson mass in general supersymmetric
  Standard Models,''
  Phys.\ Lett.\  B {\bf 302} (1993) 51
  [arXiv:hep-ph/9212305].

\bibitem{Carena:2009gx}
  M.~Carena, K.~Kong, E.~Ponton and J.~Zurita,
  ``Supersymmetric Higgs Bosons and Beyond,''
  arXiv:0909.5434 [hep-ph].
\end{thebibliography}
\end{document}